\documentclass[preprint,showpacs]{revtex4}

\usepackage[latin1]{inputenc}
\usepackage{graphicx}%
\usepackage{dcolumn}
\usepackage{amsmath}

\makeatletter
\def\btt#1{\texttt{\@backslashchar#1}}%
\DeclareRobustCommand\bblash{\btt{\@backslashchar}}%
\makeatother

\begin{document}

  \title{Theory of   self-organized traffic at light signal}

\author{Boris S. Kerner $^1$, $^2$}

\affiliation{$^1$ 
Daimler AG, RD/RTF, HPC:  G021, 71059 Sindelfingen, Germany 
}

\affiliation{$^2$  Physik von Transport und Verkehr, Universit{\"a}t Duisburg-Essen,
47048 Duisburg, Germany}



\begin{abstract}
 Based on numerical simulations of a three-phase  
  traffic flow model,  a probabilistic  theory of   traffic at the light signal is developed. 
We have found that   very complex spatiotemporal self-organized   phenomena
  determine  features of  city traffic.
 We have revealed that the breakdown   of {\it green wave} in a city
  is initiated by the emergence of   a moving synchronized flow pattern (MSP) within
  the green wave. It turns out that  a sequence of F$\rightarrow$S$\rightarrow$J transitions
  (F -- free flow, S -- synchronized flow, J -- moving queue)
  lead to traffic breakdown at the light signal.
  Both spontaneous and induced   breakdowns of the green wave have been found. 
       From a study of a variety of scenarios for  arrival traffic,
       we have found that
  there are the infinite number of capacities of traffic at the light signal, which are in a capacity range
  between a minimum capacity and maximum capacity; each of  the
  capacities   gives a flow rate at which under-saturated traffic  is
  in a metastable state with respect to the transition    to over-saturated 
  traffic. 
  The maximum capacity     depends crucially   on
  a time-dependence of the flow rate: The larger the number of vehicles  that arrive
  the light signal during the green phase, the larger the maximum capacity.
    \end{abstract}

\maketitle

 \section{Introduction \label{Int}}

  Light signals   in city intersections act as   bottlenecks  determining
  the main features of city traffic. 
 One of the basic characteristics of a well-known   Webster model~\cite{Webster} as well as
 other classical models and theories of traffic at light 
 signal
(see~\cite{Morgan,Little,Newell_1960,Newell_1965,Robertson69,Robertson79,Hunt,Michalopoulos81,Pisharody1980,Stephanopoulos79,Gartner1983,Grafton,McShane}
and   reviews~\cite{Gartner,Rakha})
  is traffic capacity at the light signal     
\begin{equation}
C_{\rm cl}=q_{\rm sat}T^{\rm (eff)}_{\rm G}/\vartheta,
\label{cl_cap}
\end{equation}
 where $q_{\rm sat}$ is the saturation flow rate, i.e., the mean flow rate from a vehicle queue
 at the light signal during green phase  when  vehicles discharge 
 from the queue to their
 maximum free speed $v_{{\rm free}}$;  $\vartheta=T_{\rm G}+T_{\rm Y}+T_{\rm R}$ is the period (cycle time) of the light signal
  that is assumed to be constant, $T_{\rm G}$,
 $T_{\rm Y}$, and $T_{\rm R}$ are   durations of the
   green,   yellow, and  red phases of the light signal, respectively;  $T^{\rm (eff)}_{\rm G}$ is
 the effective green phase time that is the portion
 of the cycle time during which vehicles are assumed to pass the light signal at constant rate $q_{\rm sat}$.
 A summary  of these and other definitions, variables, and values used   is given   in  Appendix~\ref{App_D}.

  In the classical theories (reviews~\cite{Gartner,Rakha}),   capacity $C_{\rm cl}$   (\ref{cl_cap}) 
 determines the transition from under- to over-saturated traffic.
 In under-saturated traffic,
 all vehicles, which are waiting within a  queue  during the red phase, 
 can pass the signal during the green phase. An opposite case occurs
 in over-saturated traffic and, therefore, the queue   grows.
 It is assumed~\cite{Gartner,Rakha}   that if $q_{\rm in}>C_{\rm cl}$, i.e., 
    capacity (\ref{cl_cap})  is less than the flow rate of vehicles  $q_{\rm in}$
  that arrive at the light signal
 (called  as arrival traffic rate  on the approach~\cite{Gartner}),  
 then a transition from under- to
  over-saturated traffic occurs.
 
 In the classical theories of city traffic 
  is furthermore assumed that no instabilities and no self-organization phenomena can occur in city traffic
  (reviews~\cite{Gartner,Rakha}). This is
 because  traffic lights should constitute massive deterministic
 perturbations suppressing  the  self-organized phenomena in city traffic. 
This  has also been earlier  assumed by the author    (see Sec.~22.4 in~\cite{KernerBook}
 and footnote~1 in Chap.~1 of~\cite{KernerBook2}).
In contrast, as the author has recently found, in under-saturated  traffic 
spontaneous traffic breakdown, i.e., the phase transition from under- 
to over-saturated traffic  can occur at the light signal  after a random time delay $T^{\rm (B)}$ with some
probability $P^{\rm (B)}$~\cite{Kerner2011_G}:  the  queue at the light signal 
 begins to self-grow non-reversibly leading to
 traffic gridlock.
 
 In this article, based on numerical simulations
 of a three-phase   traffic flow model we present a  probabilistic theory of  traffic at the light signal. In this   theory,
  features of city traffic are determined by
   traffic breakdown and resulting  spatiotemporal 
   self-organization traffic phenomena.
 The classical theories of city traffic 
  are   the basis for a   
  variety of light signal control systems, for example, for 
  an arterial progressive control during which vehicles
  should travel unimpeded in a city~\cite{Gartner}; this should implement
  a well-known    idea about
 a {\it green wave} in a city.  
 However,  we will reveal that
   complex self-organization traffic phenomena at the light signal 
   should be taken into account for the optimization of a green wave in  a city.
   
   The article is organized as follows. In Sec.~\ref{GW_S}, we present
   a theory of the breakdown of a green wave at an isolated light signal.
   Self-organization phenomena due to spatiotemporal interaction of the green wave with a  vehicle queue 
   are the subject of Sec.~\ref{GW_R_S}.
   In Sec.~\ref{Probability_S}, we study  
   probability of traffic breakdown.
  The  infinite number of capacities of traffic at the light signal are considered in Sec.~\ref{Capacity_S}. 
   Induced breakdown of green wave is studied in Sec.~\ref{Induced_S}. In Sec.~\ref{Diagram_S}, we study a diagram of the breakdown at
   the light signal
   and show
   that  probability of  green wave breakdown can exhibit a minimum as a function of light signal characteristics.
   Green wave breakdown occurring in a more general case of a sequence of the light signals is discussed in Sec.~\ref{Many_LS_S}. 
     A discussion of  possible applications
    of the BM (breakdown minimization)   principle
  for optimization of the green wave in a city is made in Sec.~\ref{BM_S}.
  In Sec.~\ref{Discussion_S}, we make a comparison of traffic  breakdown at highway bottleneck and bottleneck due to 
  the light signal (Sec.~\ref{High_Light_S}),     
   compare
    results of three-phase and two-phase traffic flow theories in the application to city traffic (Sec.~\ref{2Phase_S})
    as well as formulate conclusions.

  \section{Breakdown of  green wave  \label{GW_S}}

    \subsection{Model of green wave at isolated light signal \label{GW_M_S}}

    \begin{figure}
\begin{center}
\includegraphics*[width=10 cm]{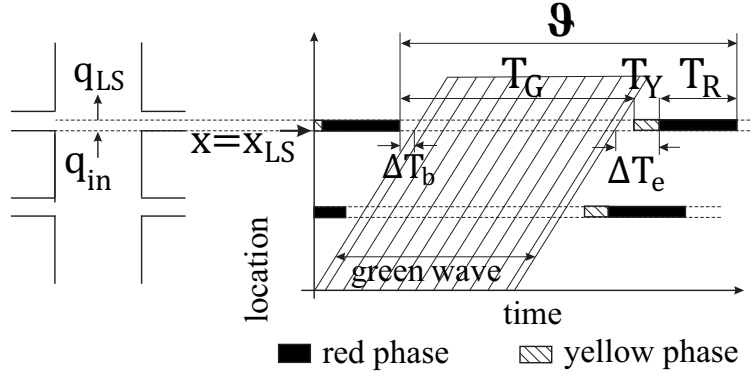}
\caption{Model of green wave.  $q_{\rm in}$ and $q_{\rm LS}$ 
 are     rates of arrival traffic at light signal and in light signal outflow, respectively.
 \label{GW_model} } 
\end{center}
\end{figure} 

In  Sec.~\ref{GW_S},
we consider   a hypothetical green wave   when 
 there is no initial vehicle queue  at the light signal.
When the green wave propagates through several identical light signals   (Fig.~\ref{GW_model}),
 probability that   spontaneous green wave breakdown occurs in {\it at least} at one of the light signals is larger than
probability $P^{\rm (B)}$ that the breakdown occurs {\it only}   at  a chosen   light signal. Therefore, firstly
to study the physics of self-organized traffic,  
we  consider the propagation of a green wave   through an {\it isolated} light signal
 at location
 $x=x_{\rm LS}$ on a single-lane city link. In this model,  $\Delta T_{\rm b}$   
 denotes a time gap between the end of the red phase   and   beginning of the  green wave; $\Delta T_{\rm e}$ 
 denotes a time gap between the end of the green wave and beginning of the red phase (Fig.~\ref{GW_model}).

Open boundary conditions  have been used in all simulations.  
   For each cycle of the light signal,   vehicles  
are    generated   
 at the road beginning $x=x_{\rm b}$  during
  a given time interval $T_{\rm GW}$ with random time headways between vehicles that 
   deviate within $10\%$ 
from a given mean gross time headway $\tau_{\rm GW}$; the latter determines
the flow rate $q_{\rm GW}=3600/\tau_{\rm GW}$ vehicles/h. The initial vehicle speed is equal to $v_{\rm free}=55$ km/h.
The time interval   between the beginning of the
   time interval $T_{\rm GW}$ and   beginning of the green phase 
   for the light signal at location $x=x_{\rm LS}$ is calculated from formula
     $((x_{\rm LS}-x_{\rm b})/v_{\rm free})-\Delta T^{\rm (ideal)}_{\rm b}$, where
    $\Delta T^{\rm (ideal)}_{\rm b}$ denotes a value $\Delta T_{\rm b}$ under
  undisturbed 
   and noiseless vehicle motion
   at the speed $v_{\rm free}$. Under such a  hypothetical vehicle motion,   
   the time gap  $\Delta T_{\rm e}=\Delta T^{\rm (ideal)}_{\rm e}=T_{\rm G}+T_{\rm Y}-T_{\rm GW}-\Delta T^{\rm (ideal)}_{\rm b}$.
  After vehicles have passed the light signal, they
leave freely the simulation area.   Even in this hypothetical model, we   reveal   the phenomenon of spontaneous breakdown of   the  
    green wave.   However, before we briefly consider
   a new feature of Kerner-Klenov microscopic three-phase traffic flow model  
   used for simulations (Sec.~\ref{Model_S}) as well as features of
   two    basic traffic localized patterns needed for the paper understanding
      (Sec.~\ref{WMJ_MSP_S}).
    
  \subsection{Three-phase microscopic stochastic traffic flow model for city traffic   \label{Model_S}}

     For a study of city traffic we have used a discrete version of the Kerner-Klenov
   stochastic three-phase microscopic model for a single-lane road whose continuum version has  initially been developed
    for highway traffic~\cite{KKl2003A,Three} that reads as follows:
    \begin{equation}
v_{n+1}=\max(0, \min(v_{{\rm free}}, \tilde v_{n+1}+\xi_{n}, v_{n}+a \tau, v_{{\rm s},n} )),
\label{final}
\end{equation}
\begin{equation}
\label{next_x}
x_{n+1}= x_{n}+v_{n+1}\tau,
\end{equation}
where 
 $n=0, 1, 2, ...$ is number of time steps,
$\tau=1$ s is a time step~\cite{Time_Step}, $x_{n}$ and $v_{n}$ are the vehicle coordinate and speed
at time step $n$,   $a$ is the maximum acceleration, $v_{\rm free}$ is a maximum speed in free flow,
$\tilde v_{n}$ is the vehicle speed  without  speed fluctuations $\xi_{n}$,
$v_{{\rm s}, n}$ is a safe speed. 
   
   In addition to a lower   speed   $v_{\rm free}$~\cite{Kerner2011_G}, in city traffic we should ensure
   a larger  vehicle acceleration from a standstill in a queue 
      in comparison with a relatively small   
     acceleration $a$ in (\ref{final}) chosen in accordance with empirical features of  
   a phase transition from free flow to synchronized flow (F$\rightarrow$S transition)~\cite{KKl2003A,Kerner2011_G}.
   This larger acceleration  is required to satisfy an empirical value of   
    lost time during the green phase $\delta t=T_{\rm G}-T^{\rm (eff)}_{\rm G}\approx$ 3--4 s~\cite{Gartner,Rakha}.
    We have made the following  model
   development. When the speed difference $\Delta v_{n} =v_{\ell, n}-v_{n}$ between the vehicle speed $v_{n}$ and  speed of the preceding vehicle
   $v_{\ell,n}$ is great enough and/or the acceleration of the preceding vehicle $a_{\ell,n}$ is large enough satisfying  condition
   \begin{equation}
   \Delta v_{ n} + a_{ \ell,n}\tau \geq \Delta v_{\rm a},
   \label{jump}
   \end{equation}
    then rather than  acceleration $a$, the larger maximum acceleration $k_{\rm a}a$ with $k_{\rm a}>1$ is used;
   in (\ref{jump}), $\Delta v_{\rm a}$ is constant.
   Otherwise, the maximum acceleration remains to be equal to $a$ of the original model~\cite{Kerner2011_G,KKl2003A}. 
Because all other model functions are the same as those in the  Kerner-Klenov model 
for  a single-lane road~\cite{KKl2009,KKl2010}, the  functions and  parameters are given in Appendix~\ref{App1}.
 The physical sense
   of condition (\ref{jump}) is as follows. If    (\ref{jump}) is not satisfied,  rules of vehicle motion 
   are  the same
   as those of the initial model~\cite{KKl2003A,Kerner2011_G}.
    However, when
  condition (\ref{jump}) is satisfied, rather than 
    car-following within   synchronized flow at a small acceleration $a$,  the vehicle follows    
  the preceding vehicle  with a   greater acceleration
    $k_{\rm a}a$~\cite{Ac_Fol}. As a result, the model time lost $\delta t=T_{\rm G}-T^{\rm (eff)}_{\rm G}\approx$
   3.2 s satisfies empirical values.
   
 As in~\cite{Kerner2011_G}, in the model
vehicles decelerate at the upstream front of a queue at the light signal as they do this at the upstream front
of a wide moving jam  propagating on a road without light signals~\cite{KernerBook,KernerBook2}. 
During the green phase, vehicles accelerate at the downstream queue front (queue discharge) with a random
time delay as they do it at the downstream jam front; in other words, the well-known saturation flow rate of queue discharge is equal to
the jam outflow $q_{\rm out}$ under the condition that vehicles accelerate to the maximum speed  $v_{\rm free}$, i.e.,
in this case $q_{\rm sat}=q_{\rm out}$, which is equal to 1808 vehicles/h under chosen model parameters.
During the yellow phase 
 the vehicle  passes the light signal location, if the vehicle can do it until 
 the end of the yellow phase; otherwise, the vehicle 
  comes to a stop at the light signal. 
  
  \subsection{Two basic moving localized patterns in three-phase theory of city traffic   \label{WMJ_MSP_S}}

\begin{figure}
\begin{center}
\includegraphics*[width=10 cm]{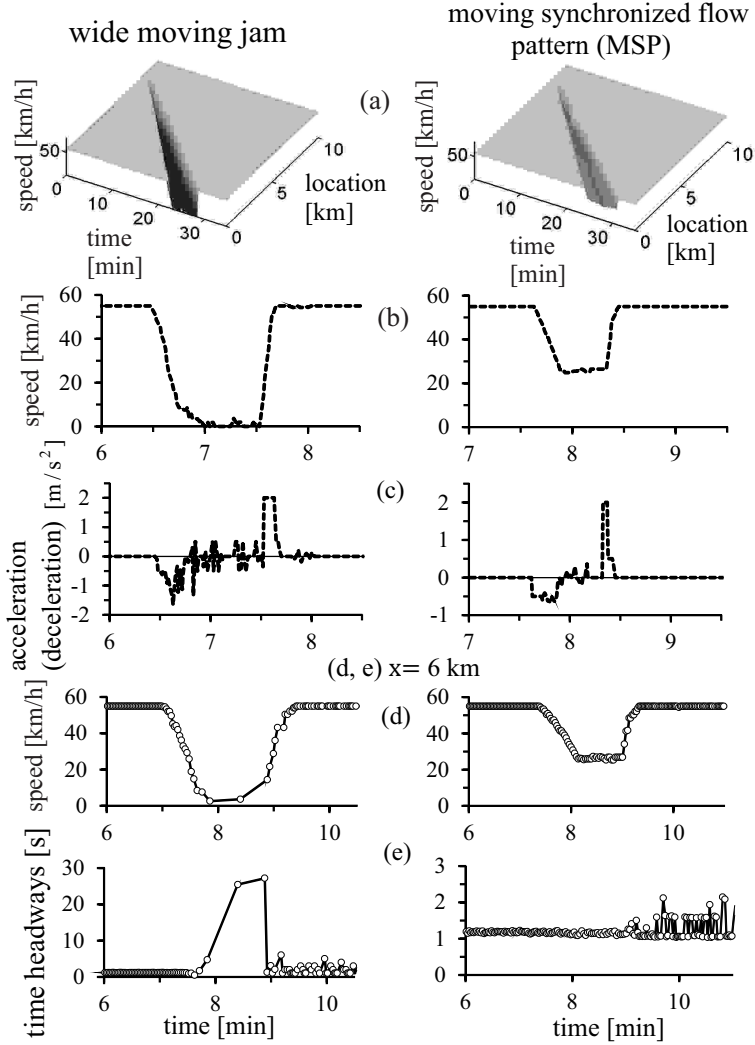}
\caption{Simulations of wide moving jams (left panel) and moving synchronized flow pattern (MSP) (right panel)
on homogeneous single-lane road without bottlenecks:
(a) Speed in time and space.
(b, c) Microscopic speed (b) and acceleration (deceleration) (c) for a vehicle moving through
the jam (left) or MSP (right). (d, e) Microscopic speed (d) and time headways (e) of vehicles measured at a virtual detector
at   location $x=$ 6 km.
  \label{WMJ_MSP} } 
\end{center}
\end{figure}
  
  As in highway traffic~\cite{KernerBook,KKl2003A,KKl2009,KKl2010},  
  there are two    qualitatively different 
  localized patterns which play the basic role in  theory of city traffic:
  a wide moving jam (Fig.~\ref{WMJ_MSP}, left panel) 
  and a moving synchronized flow pattern (MSP) (Fig.~\ref{WMJ_MSP}, right panel). The wide moving jam
  satisfies the microscopic criterion for the wide moving jam phase~\cite{KKH,KKHR,KernerBook2}:  there is a flow interruption interval within
  the jam, i.e., a long
  time headway(s) between vehicles (Fig.~\ref{WMJ_MSP} (e), left panel) that is considerably longer than the mean time delay
  of vehicle acceleration from a standstill within the jam. During
  the green phase,
  specifically, after the queue discharge flow increases to $q_{\rm sat}$, the moving queue and   wide moving jam   exhibit the same
   features; 
  therefore, the moving queue can be considered the wide moving jam phase (J) of congested traffic   in a city.
  
  In contrast with the moving queue,
   there is no flow interruption   within the MSP (Fig.~\ref{WMJ_MSP} (e), right panel) -- the microscopic criterion for
  the jam does not satisfy, i.e., the MSP  belongs to the synchronized flow phase (S).
  To induce an MSP   in  free flow, a local disturbance should exceed a critical value $\Delta v^{\rm (cr)}_{\rm FS}$  
  required for an
  F$\rightarrow$S transition. Respectively, to induce a
  moving queue
   in   free flow a local disturbance should exceed another critical value $\Delta v^{\rm (cr)}_{\rm FJ}$ required for an
  F$\rightarrow$J transition. 
  However, at each given flow rate, at which
  either an F$\rightarrow$S or F$\rightarrow$J transition is possible,
  $\Delta v^{\rm (cr)}_{\rm FS}\ll
  \Delta v^{\rm (cr)}_{\rm FJ}$. 
  This means that there is a wide range of speed disturbance amplitudes $\Delta v_{\rm dis}$ satisfying condition
  $\Delta v^{\rm (cr)}_{\rm FS}\leq \Delta v_{\rm dis} < \Delta v^{\rm (cr)}_{\rm FJ}$
  within which no  moving queues can emerge, whereas
  MSP does occur in free flow.

\subsection{Emergence   of moving synchronized flow pattern (MSP) within green wave \label{MSP_S_em}}

     \begin{figure}
\begin{center}
\includegraphics*[width=8 cm]{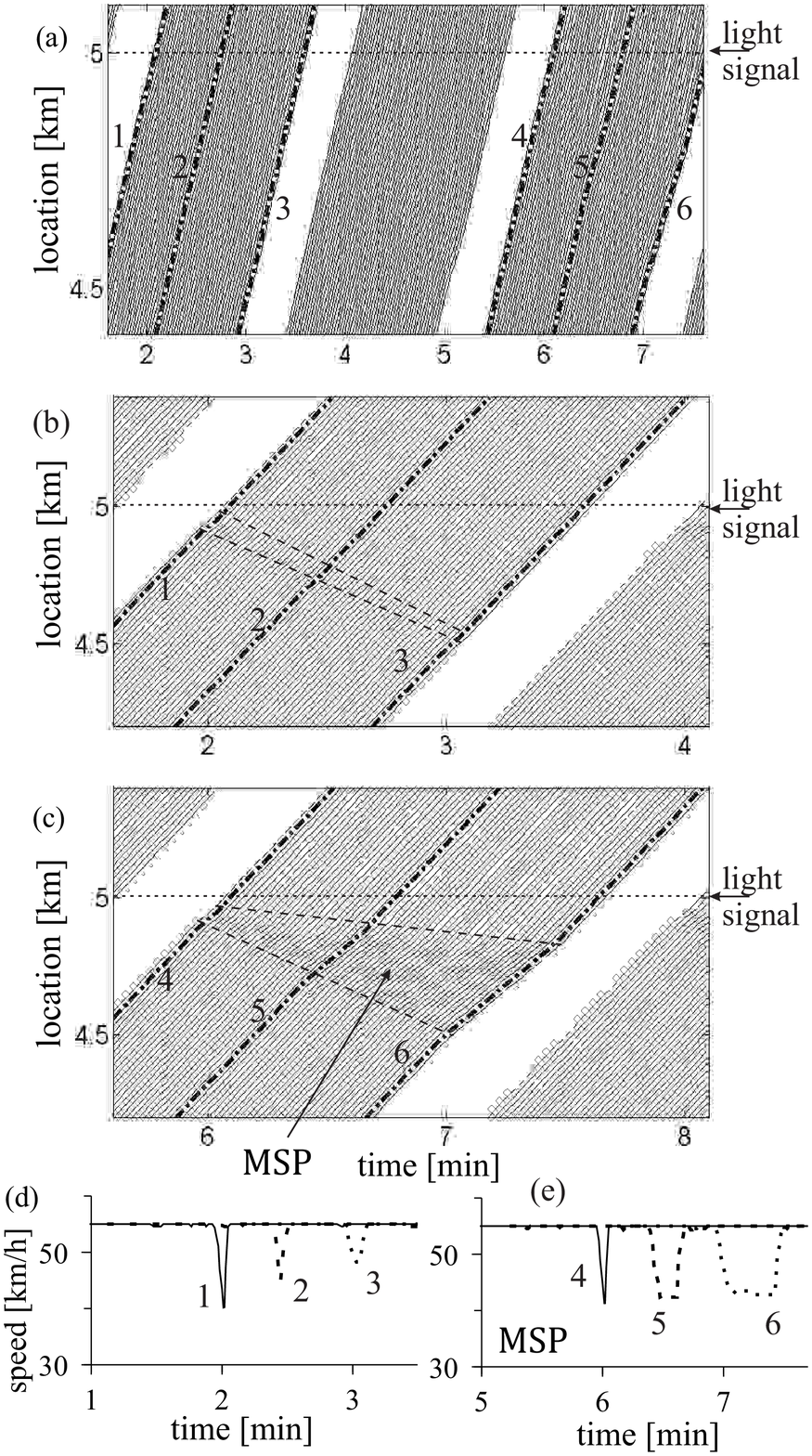}
\caption{Emergence of MSP within green wave at $\vartheta=$ 120 s, $T_{\rm R}=$ 20 s,  $T_{\rm Y}=$ 2 s,
  $\Delta T^{\rm (ideal)}_{\rm b}= $ 3 s, and   $T_{\rm GW}=90$ s: 
  (a--c) Vehicle trajectories
within green wave. (d, e) Microscopic speeds of vehicles
moving through MSPs whose numbers are related to trajectories in (a--c), respectively.
In (b, c), disturbance and MSP are marked by dashed curves. $q_{\rm GW}= 2316$ vehicles/h~\cite{Long}.
 \label{GW_3_2316_MSP} } 
\end{center}
\end{figure}
 
 At the first glance,    all green waves propagate undisturbed   over different cycles of the light signal 
 (Fig.~\ref{GW_3_2316_MSP} (a)).
 However, if we consider   vehicle trajectories in a larger scale (Figs.~\ref{GW_3_2316_MSP} (b, c)), we find
 that there is a small speed disturbance  at the beginning of each green wave. To understand this, we note that
 if a driver sees the red phase, then to stop at the light signal location $x=x_{\rm LS}$
   she/he should begin to decelerate   at some distance $\Delta x_{\rm dis}$
 from the light signal. When the driver is at location $x=x_{\rm LS}-\Delta x_{\rm dis}$
  and she/he moves at the   speed $v_{\rm free}$, 
 it takes the driver a time $\Delta T_{\rm  dis}$
  to reach the light signal; $\Delta T_{\rm  dis}\approx$ 7 s at chosen model parameters.
  Thus when $\Delta T_{\rm  dis}-\Delta T^{\rm (ideal)}_{\rm b}>0$  (Fig.~\ref{GW_3_2316_MSP}),  
  the driver
  reaching location  $x=x_{\rm LS}-\Delta x_{\rm dis}$   decelerates during
   the time interval $\Delta T_{\rm  dis}-\Delta T^{\rm (ideal)}_{\rm b}$, while seeing   the red phase. 
 After the green phase appears, the driver accelerates to the maximum speed $v_{\rm free}$.
This explains  speed
disturbance occurrence (curves 1 and 4 in Figs.~\ref{GW_3_2316_MSP} (d, e)). 
 
 In Fig.~\ref{GW_3_2316_MSP}, we have chosen   
 green wave parameters at which the initial disturbance amplitude is close to a critical one:
 In some of the light signal cycles, 
 speed  disturbances are smaller than   the critical disturbance; therefore, no MSP occurs while   disturbances dissolve 
  (trajectories 1--3 in Figs.~\ref{GW_3_2316_MSP} (b, d)).
 In other cycles,  speed  disturbances are larger than   the critical disturbance
 with resulting MSP emergence
   (trajectories 4--6 in Figs.~\ref{GW_3_2316_MSP} (c, e)).  
Any MSP  and any disturbance   fully disappears at the end of each green wave and, therefore, in Fig.~\ref{GW_3_2316_MSP}
the random  process
 of the disturbance occurrence  and development within a subsequent green wave is independent on the former green wave~\cite{qGW_cr}.

  Through the MSP emergence 
 time gaps  $\Delta T_{\rm b} $ and $\Delta T_{\rm e} $ (Fig.~\ref{GW_model}) are
respectively longer and shorter than  $\Delta T^{\rm (ideal)}_{\rm b} $ and $\Delta T^{\rm (ideal)}_{\rm e} $ calculated
for an undisturbed green wave (Sec.~\ref{GW_M_S}).
Even in the same simulation realization (run)~\cite{Realization},
  parameters of  MSPs  that occur in different cycles  
  are random values. Consequently, within time interval $0<t<34$ min the gaps    
$\Delta T_{\rm b} $ and $\Delta T_{\rm e} $ change randomly 
for different green waves 
between 3.8--4.39 s and 0.07--5.15 s, respectively; the
mean  values of   $\Delta T_{\rm b} $ and $\Delta T_{\rm e} $ are respectively 3.96 s and 2.38 s  
  (compare with
 $\Delta T^{\rm (ideal)}_{\rm b}=3 $ s and $\Delta T^{\rm (ideal)}_{\rm e}=7 $ s used in Fig.~\ref{GW_3_2316_MSP}).

\subsection{Common stages of green wave breakdown:
Features of F$\rightarrow$S$\rightarrow$J transitions \label{Br_S}}

Although an MSP emerges spontaneously in some of the light signal cycles   
(Figs.~\ref{GW_3_2316_MSP} (c, e)), no breakdown   have been observed up to   $t=$ 34 min. However, 
if we consider the simulation realization shown in Fig.~\ref{GW_3_2316_MSP}
at a longer time, we do find the phenomenon of the green wave breakdown (Figs.~\ref{GW_3_2316} and~\ref{GW_3_Break}).
We have found  that
the phenomenon of  green wave breakdown      
begins  randomly   and it can be considered consisting of
 the following   stages:
 
        \begin{figure}
\begin{center}
\includegraphics*[width=9 cm]{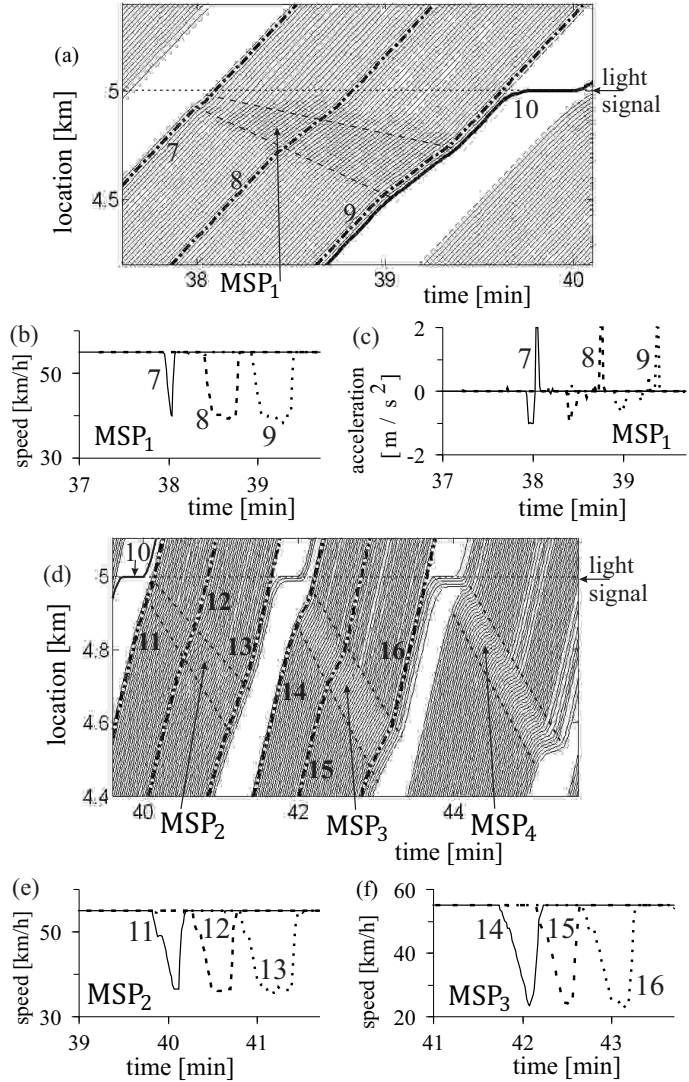}
\caption{Breakdown of green wave in simulation realization shown in   Fig.~\ref{GW_3_2316_MSP}: (a, d) Vehicle trajectories in different scales
in which
 MSPs are marked by dashed curves. (b, c, e, f) Microscopic speeds (b, e, f) and acceleration (deceleration) (c) of vehicles
moving thorough MSPs propagating through different green waves.
Curves 7--16 in (b, c, e, f) are related to corresponding vehicle trajectories marked in (a, d).
  \label{GW_3_2316} } 
\end{center}
\end{figure}

    \begin{figure}
\begin{center}
\includegraphics*[width=10 cm]{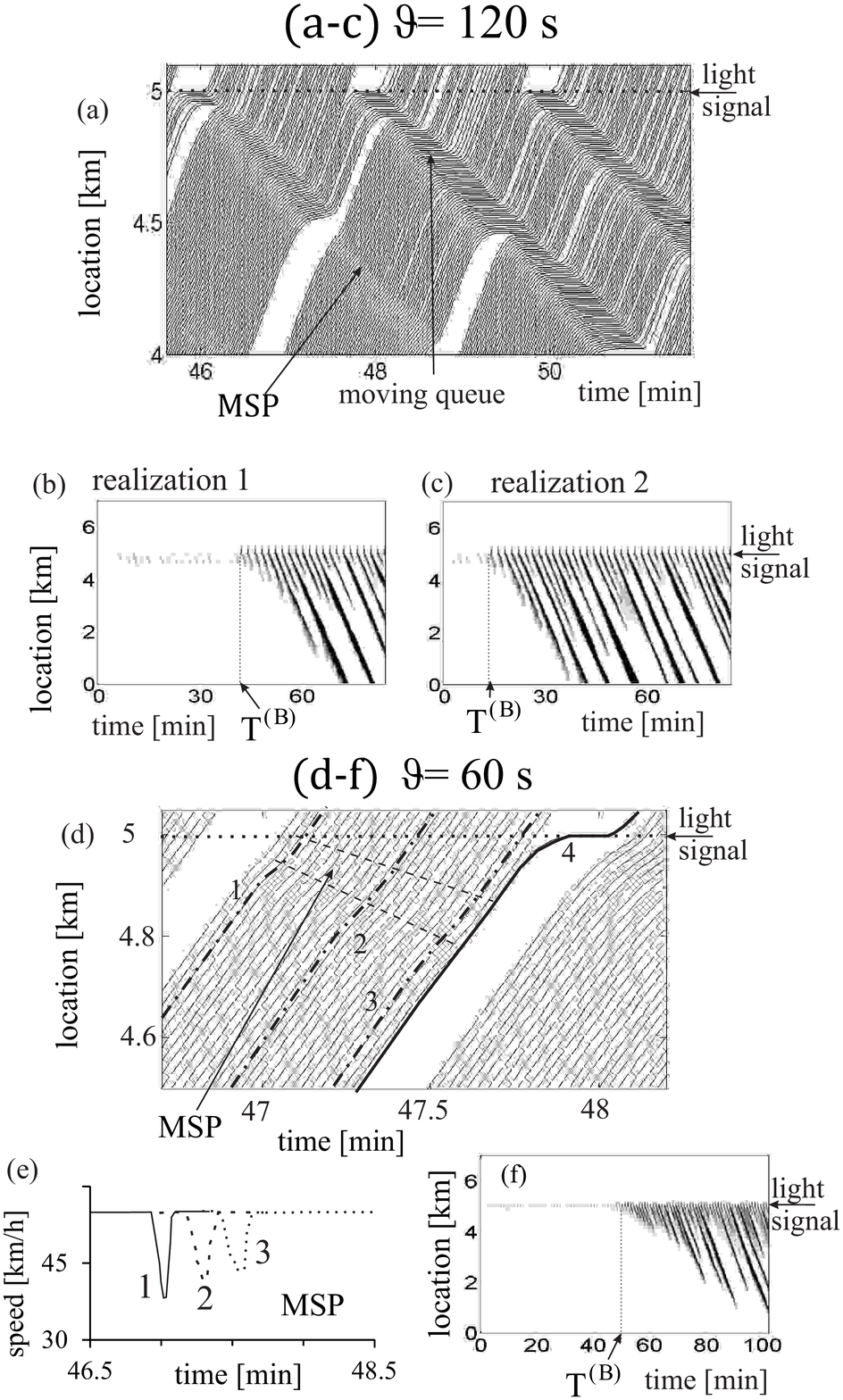}
\caption{Characteristics of 
breakdown of green waves with
$\Delta T^{\rm (ideal)}_{\rm b}=3 $ s for cycle times $\vartheta=120$ s (a--c) and $\vartheta=60$ s (d--f): 
(a, d) Vehicle trajectories.  (b, c, f)
Speed   data  in space-time plane 
presented  by regions with variable shades of gray for two 
different realizations 1 (b) and 2 (c)~\cite{Realization}. (e) Microscopic speeds 
 of vehicles 1--3
moving through MSP in (d)  marked by dashed curves. 
In (a, b), the same realization 1 as that in Figs.~\ref{GW_3_2316_MSP} 
and~\ref{GW_3_2316} is used. In (a--c),
model parameters are the same as those 
in Fig.~\ref{GW_3_2316_MSP}. In (d--f), $\vartheta=$ 60 s, $T_{\rm R}=$ 10 s, $T_{\rm Y}=$ 2 s, $T_{\rm GW}=45$
s, $q_{\rm GW}=$   2110 vehicles/h~\cite{Long}.
  \label{GW_3_Break} } 
\end{center}
\end{figure}

(i)   {\it  An MSP occurs and propagates upstream within the green wave} (MSP labeled by $\lq\lq$$MSP_{1}$" 
in Figs.~\ref{GW_3_2316} (a--c)). 

(ii)    {\it The last vehicle or a few of the last  vehicles    of the green wave come to a standstill:
The random process of the green wave breakdown begins }
(bold   trajectory 10  
in Fig.~\ref{GW_3_2316} (a)). 
The physics of stage (ii) is as follows:      
Vehicles exhibit delays    moving through the MSP.  When
the    delay of the last vehicle of the green wave  becomes randomly 
longer than $\Delta T^{\rm (ideal)}_{\rm e}$, the vehicle must stop at the following red phase. 
 The random nature of this vehicle stop is associated with 
  random characteristics of a disturbance and resulting   MSP.
In some other simulation realizations~\cite{Realization},
 rather than only the last vehicle of the green wave (Fig.~\ref{GW_3_2316} (a)), 
 two or more vehicles must stop at the light signal.
 
(iii) {\it Synchronized flow speeds in MSPs occurring within the subsequent green waves
    decreases.}
The    vehicle(s) stopped at the light signal (item (ii)) passes it during the
next green phase. This  forces
vehicles  of the following green wave to decelerate stronger
introducing
a larger disturbance within the green wave (trajectory 11 in Fig.~\ref{GW_3_2316} (d, e)) than
in the signal cycles    shown in Figs.~\ref{GW_3_2316_MSP}  and~\ref{GW_3_2316} (a, b).
This result is an MSP (labeled by $MSP_{2}$    in Figs.~\ref{GW_3_2316} (d, e)) 
with   lower  speeds (trajectories 12 and 13 in Fig.~\ref{GW_3_2316} (d, e)).
Consequently, a larger number of vehicles at the end of the green wave exhibit a  longer delay than $\Delta T^{\rm (ideal)}_{\rm e}$. 
Therefore,
more vehicles must stop
at the following red phase  ($t\approx $ 42 min in Fig.~\ref{GW_3_2316} (d)). The discharge of this longer vehicle queue 
  at the next green phase takes a longer time.  This increases 
  further the disturbance amplitude    at the beginning of the following green wave
  with the further decrease in the speed within  the emergent MSP  ($MSP_{3}$  
   in Figs.~\ref{GW_3_2316} (d, f)).
This results in the subsequent increase in the number of vehicles that must stop at the light signal: five vehicles have stopped
at $t\approx $ 44 min in Fig.~\ref{GW_3_2316} (d). The discharge of these vehicles causes MSP emergence 
($MSP_{4}$ in Fig.~\ref{GW_3_2316} (d)) with a very low synchronized flow speed, and so on.

(iv) {\it The breakdown of the green wave occurs randomly with destroying of the green wave
leading to the appearance of over-saturated traffic.} 
Stage (iii)) ends abruptly at some of the cycles of the light signal:
Instead of an MSP, at the beginning of the next green wave
 a moving queue     appears that propagates through  the  green wave
 (moving queue in Fig.~\ref{GW_3_Break} (a)):
 The green wave breakdown has occurred.
 After the breakdown has occurred, the queue length at the light signal grows, i.e., over-saturated traffic occurs. 
The breakdown   occurs when vehicles stopped at the beginning of the red phase  
  forms a critical   queue:
  When the critical queue has been reached, 
    vehicles  of the next green wave must stop approaching the end of this queue.
    The cycle   
   at which the critical queue has been formed determines a time delay
  of the breakdown denoted by $T^ {\rm (B)}$ (Fig.~\ref{GW_3_Break} (b)).
          $T^{\rm (B)}$ 
 is a  random value, which can change considerably
 in different simulation realizations~\cite{Realization}  
  (Fig.~\ref{GW_3_Break} (b, c)).
   
The MSP emergence within the green wave is associated with an F$\rightarrow$S  transition.
The   transformation of the MSP into a moving queue  (stages (iii) and (iv))
can be considered an S$\rightarrow$J transition. Thus
the green wave breakdown is associated with   a  sequence of
F$\rightarrow$S$\rightarrow$J transitions.
In these F$\rightarrow$S$\rightarrow$J transitions, both 
 an F$\rightarrow$S transition and S$\rightarrow$J transition
are   random events.
A     time interval between random 
time instants of the F$\rightarrow$S
and  S$\rightarrow$J transitions   can be   much longer than   the light signal cycle.
 We have found that these qualitative features of the green wave breakdown  remain for a broad range of   light signal parameters.
 This  is illustrated in Fig.~\ref{GW_3_Break} (d--f) for $\vartheta=60$ s.  

Simulations show that when at a given $q_{\rm GW}$ the value $\Delta T^{\rm (ideal)}_{\rm b}$     decreases,
the disturbance amplitude within the green wave  increases resulting in an MSP with a low speed.
 In general, 
   the larger the  flow rate $q_{\rm GW}$ and/or the shorter the value $\Delta T^{\rm (ideal)}_{\rm b}$, the lower the   speed
 within MSPs.
 At some chosen $q_{\rm GW}$ and $\Delta T^{\rm (ideal)}_{\rm b}$, it can turn out that
 rather than the   MSP   propagates to the end of the green wave, green wave breakdown occurs during MSP propagation through the
 green wave due to  MSP
    transformation     into a moving queue.

\subsection{Green wave breakdown caused by growing speed disturbances along green wave \label{Green_along_S}}
 
\begin{figure}
\begin{center}
\includegraphics*[width=8 cm]{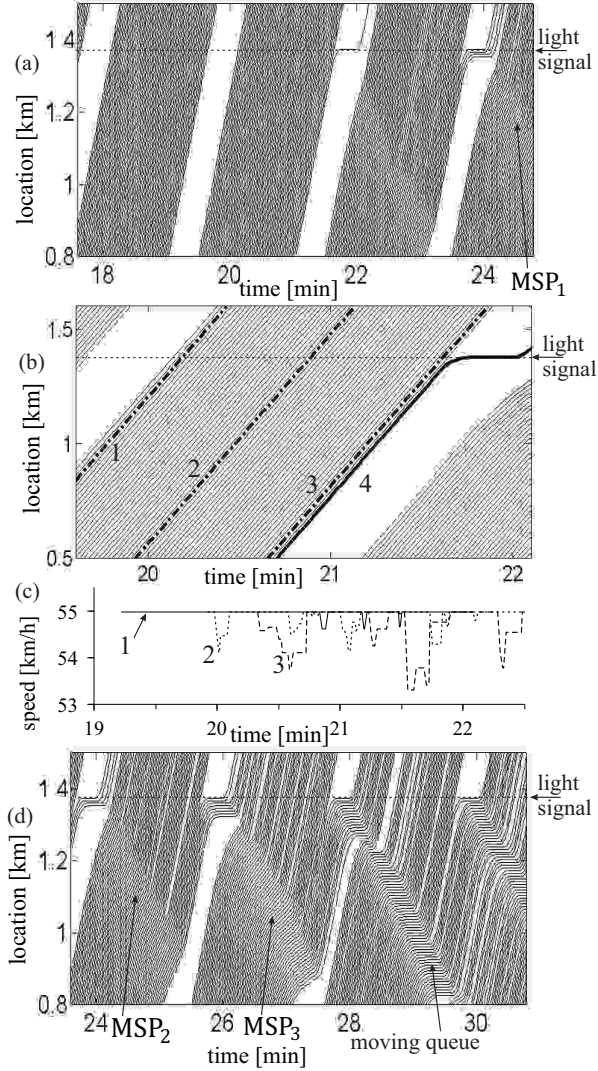}
\caption{Green wave breakdown    under condition (\ref{dist_non}):
  (a, b, d)  Vehicle trajectories.
(c) Microscopic speeds 
 of vehicles whose numbers related to trajectories shown in (b), respectively.  
($\Delta T^{\rm (ideal)}_{\rm b}, \ x_{\rm LS}-x_{\rm b}, \ q_{\rm GW})=$  
 (8, 1375, 2382) (s, m, vehicles/h)~\cite{Long}.
Other model parameters are the same as those in Fig.~\ref{GW_3_2316_MSP}.
For different green waves occurring within time interval $0<t<19$ min  values
$\Delta T_{\rm b} $ and $\Delta T_{\rm e} $ (Fig.~\ref{GW_model}) 
change between 7.89--8.16 s and 0.06--0.88 s, respectively.
\label{GW_8_2278} } 
\end{center}
\end{figure}

When  at a given   $q_{\rm GW}$ the value $\Delta T^{\rm (ideal)}_{\rm b}$ increases,
the disturbance amplitude at the green wave beginning  
decreases. 
 Under condition
 \begin{equation}
\Delta T_{\rm  dis}-\Delta T^{\rm (ideal)}_{\rm b}\leq 0,
\label{dist_non}
\end{equation} 
in   contrast with   green waves considered in Sec.~\ref{Br_S}  no
speed disturbance   appears at the beginning of the green wave.
However, we have revealed that even in this case the random time-delayed breakdown of the green wave can occur  
with  probability $P^{\rm (B)}>0$
  during a chosen  time interval $T_{\rm ob}$ (Fig.~\ref{GW_8_2278}).
 
The physics of this phenomenon is associated with
many small local speed disturbances {\it along} the green wave. They  begin to grow
at different road locations along the green wave when
 the flow rate  $q_{\rm GW}$   is great enough~\cite{Dis_Crit}.  The longer   
 the road length   $x_{\rm LS}-x_{\rm b}$ (Fig.~\ref{GW_model}), the more probable that the speed disturbances become large enough
    before the green wave reaches the light signal; therefore, simulations show that  the shorter $x_{\rm LS}-x_{\rm b}$,
    the larger $q_{\rm GW}$ at which  the breakdown occurs   with the same probability.

\section{Self-organization phenomena due to spatiotemporal interaction of green wave with  queue at light signal \label{GW_R_S}}

\begin{figure}
\begin{center}
\includegraphics*[width=9 cm]{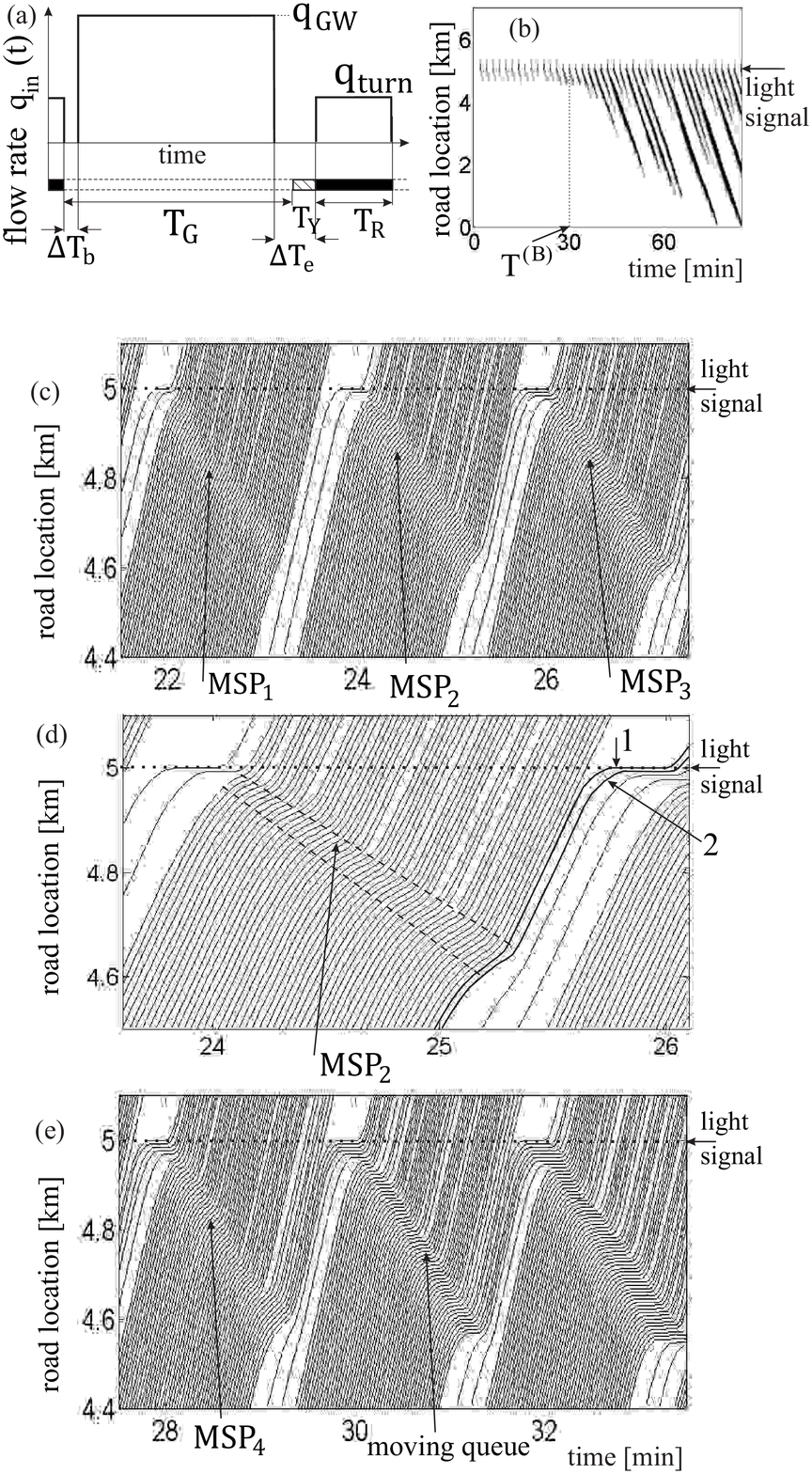}
\caption{Breakdown of green wave through
turning-in traffic: (a) The time-dependence of the  flow rate   
 $q_{\rm in}(t)$.
(b) Speed data  presented by regions   with variable shades of gray. 
(c--e) Vehicle trajectories.
$q_{\rm GW}=$ 2000    and  $q_{\rm turn}=$   400 vehicles/h. As
the flow rate $q_{\rm GW}$ (Sec.~\ref{GW_M_S}), $q_{\rm turn}$ is determined through
a given mean gross time headways $\tau_{\rm turn}$: $q_{\rm turn}=3600/\tau_{\rm turn}$ vehicles/h.
Other model parameters are the same as those in Fig.~\ref{GW_3_2316_MSP}.  
  \label{Hom_GW_turn_2000_400_3} } 
\end{center}
\end{figure}

Hypothetical green waves discussed in Sec.~\ref{GW_S}
are a rough simplification of    traffic at the light signal.
In reality, there is usually 
  turning-in traffic, which  refers
to traffic from the cross street that enters the lane on which the
green wave travels. 
Turning-in traffic leads to a queue build during the red phase. The discharge of this queue   can effect on the green wave considerably.
We simulate turning-in traffic   through flow with a rate $q_{\rm turn}$
      occurring during the red phase; we assume that $q_{\rm turn}<q_{\rm GW}$ 
      (Fig.~\ref{Hom_GW_turn_2000_400_3} (a)).

When  $q_{\rm turn}$ is not large (Fig.~\ref{Hom_GW_turn_2000_400_3} (b--e)), stages (i)--(iv) of the green wave breakdown
 are qualitatively the same
    as those for $q_{\rm turn}=0$ (Sec.~\ref{Br_S}):
The queue discharge  causes a speed disturbance at the beginning of the green wave
 with MSP emergence ($MSP_{1}$ and $MSP_{2}$  in 
Fig.~\ref{Hom_GW_turn_2000_400_3} (c, d)) (stage (i) of Sec.~\ref{Br_S}). Through vehicle delays within an MSP,
 after a random time  interval the vehicle queue build during the red phase    increases   
in comparison with   the initial queue caused by the flow rate $q_{\rm turn}$.
This queue increase occurs because one or several last vehicles at the end of the green wave have to stop at the light signal (stage (ii)) 
(Fig.~\ref{Hom_GW_turn_2000_400_3} (d), where      the stopped vehicles of the green wave
are related to bold trajectories 1 and 2, i.e.,
the queue increases from two vehicles associated with
 turning-in traffic  to four vehicles). The speed within the emergent MSP decreases ($MSP_{3}$ and $MSP_{4}$  in 
Fig.~\ref{Hom_GW_turn_2000_400_3} (c, e)) (stage (iii)). After a random time interval,
  the queue growth results in the breakdown: a moving queue is formed 
(moving queue in 
Fig.~\ref{Hom_GW_turn_2000_400_3} (e)) (stage (iv)).

     \begin{figure}
\begin{center}
\includegraphics*[width=9 cm]{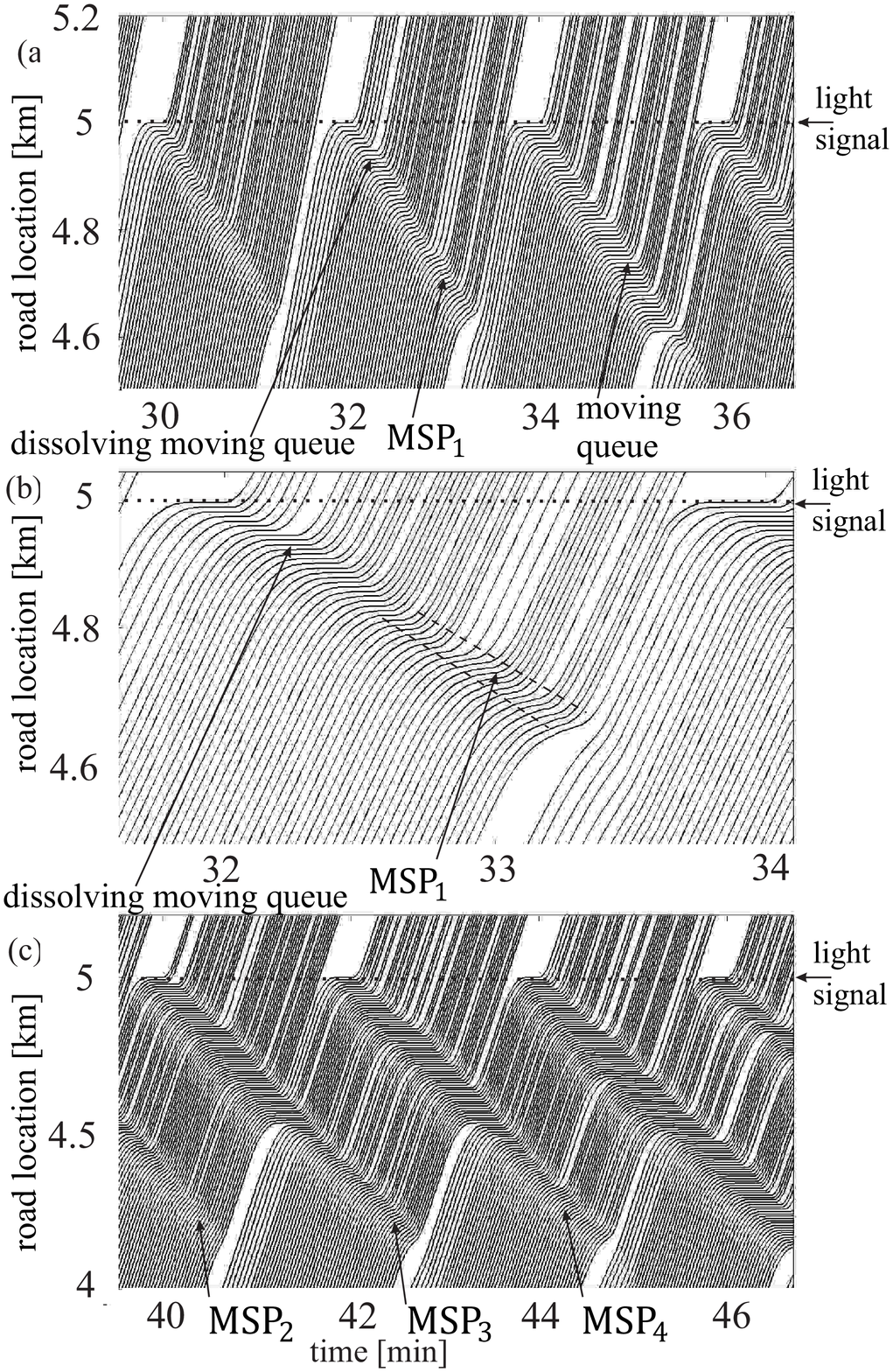}
\caption{Breakdown of green wave through
turning-in traffic simulated with  
 $q_{\rm in}(t)$ shown in Fig.~\ref{Hom_GW_turn_2000_400_3} (a): 
 Vehicle trajectories. 
$q_{\rm GW}=$ 1714    and  $q_{\rm turn}=$   1200 vehicles/h.
Other model parameters are the same as those in Fig.~\ref{GW_3_2316_MSP}.   
  \label{Hom_GW_turn_1714_1200_3} } 
\end{center}
\end{figure}

If   $q_{\rm turn}$ increases, then with the same
probability traffic breakdown occurs  at smaller flow rate $q_{\rm GW}$. In general,
qualitative phenomena of MSP emergence within the green wave with the subsequent random green wave breakdown remain
the same as those presented above.
However, when  $q_{\rm turn}$ is large enough,
the  queue  cannot fully dissolve before
 the green wave reaches the light signal (Fig.~\ref{Hom_GW_turn_1714_1200_3} (a)).
When $q_{\rm GW}<q_{\rm sat}$ (Fig.~\ref{Hom_GW_turn_1714_1200_3}),
 the queue dissolves during its propagation through the green wave while transforming into an MSP
(dissolving moving queue
and $MSP_{1}$    in Fig.~\ref{Hom_GW_turn_1714_1200_3} (a, b)). 
The flow rate  $q_{\rm GW}$  in  Fig.~\ref{Hom_GW_turn_1714_1200_3} is 
smaller than the threshold flow rate for MSP existence~\cite{KernerBook}. Therefore, the  MSP begins also to dissolve during
its propagation within the green wave. Nevertheless, it takes a relatively long time for
this MSP dissolution:   
Vehicle delays become long enough for the increase in number of vehicles that stop at the light signal  resulting in
the breakdown  (Fig.~\ref{Hom_GW_turn_1714_1200_3} (a)).

\begin{figure}
\begin{center}
\includegraphics*[width=9 cm]{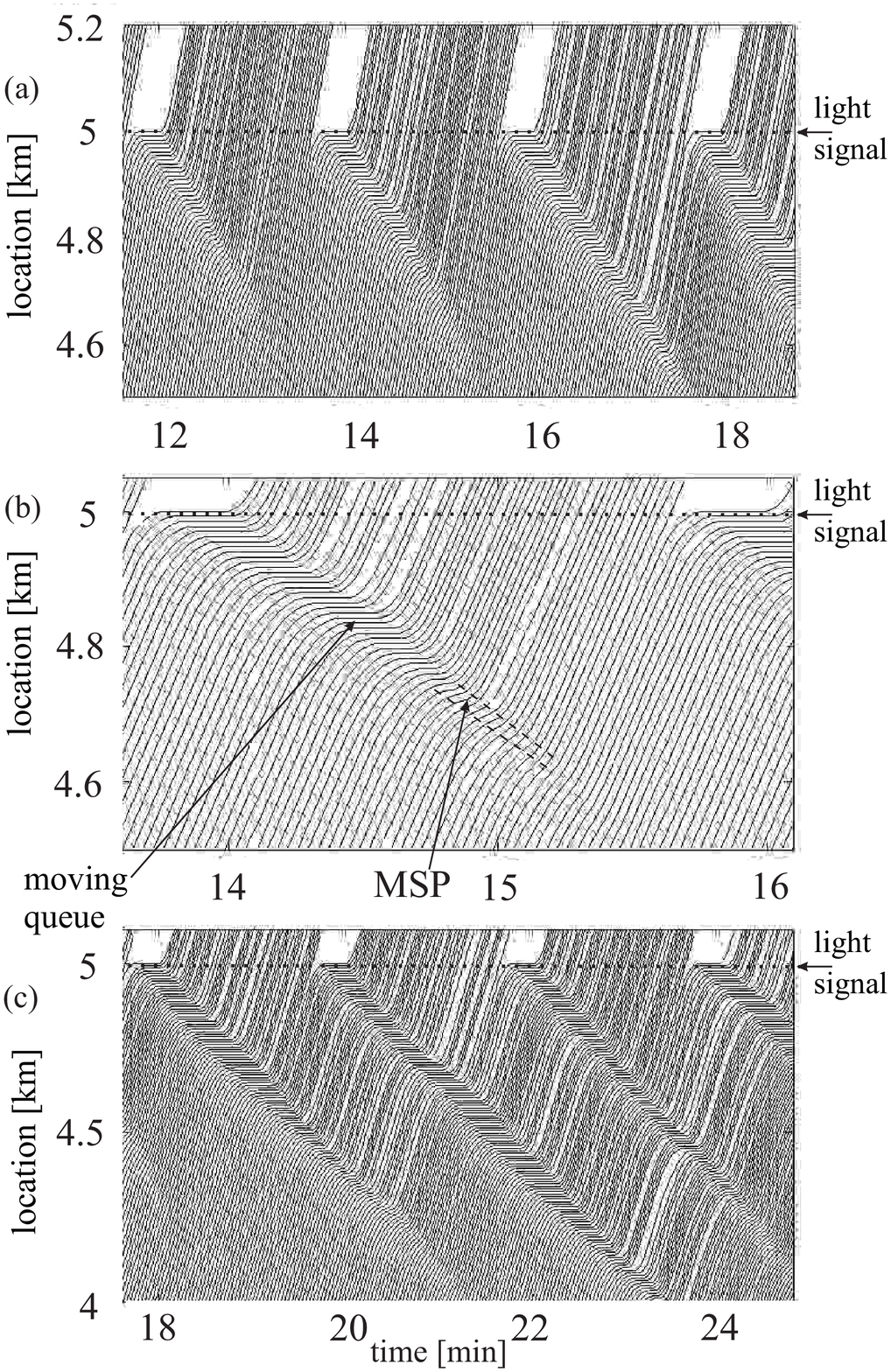}
\caption{Breakdown   at light signal at time-independent   
 $q_{\rm in}$:  Vehicle trajectories. $q_{\rm in}=$ 1565 vehicles/h.
Other model parameters are the same as those in Fig.~\ref{GW_3_2316_MSP}.   
  \label{Hom} } 
\end{center}
\end{figure}

The   phenomena presented in Figs.~\ref{Hom_GW_turn_2000_400_3}
and~\ref{Hom_GW_turn_1714_1200_3} remain qualitatively for any chosen difference   $q_{\rm GW}-q_{\rm turn}>0$ when  
the flow rates $q_{\rm turn}$ and  $q_{\rm GW}$ are chosen on the way that
  probability of traffic breakdown does not change considerably ($P^{\rm (B)}\approx$ 0.8
  in Figs.~\ref{Hom_GW_turn_2000_400_3} and~\ref{Hom_GW_turn_1714_1200_3}).
  However,  the larger $q_{\rm turn}$ and the smaller $q_{\rm GW}$,
 the longer the queue dissolution within the green wave and, therefore,
 the shorter the time interval for 
  MSP propagation within the green wave. However, even in a limit case of a time-independent
  flow rate $q_{\rm in}$~\cite{Kerner2011_G} 
 we have found MSP emergence    at the end of the green phase.
 This MSP emergence does govern the time delayed
  breakdown at the light signal. 
 One of the general results of the study made above   is that a   complex  time-sequence of
F$\rightarrow$S$\rightarrow$J transitions is responsible for the breakdown phenomenon at the light signal (Fig.~\ref{Hom}).

 After green wave breakdown has occurred, the resulting
   dynamics of  over-saturated traffic upstream of the light signal
 exhibits  complex spatiotemporal   coexistence
moving queues  and MSPs. For example, MSPs   can result from dissolving queues
($MSP_{2}$--$MSP_{4}$  resulting from dissolving queues  in Fig.~\ref{Hom_GW_turn_1714_1200_3} (c)) or an MSP can
emerge at the beginning of the green wave  at a relatively long distance 
  upstream of the queue
  (MSP  in Fig.~\ref{GW_3_Break} (a)).

   \section{Probability of traffic breakdown at light signal \label{Probability_S}}

\begin{figure}
\begin{center}
\includegraphics*[width=10 cm]{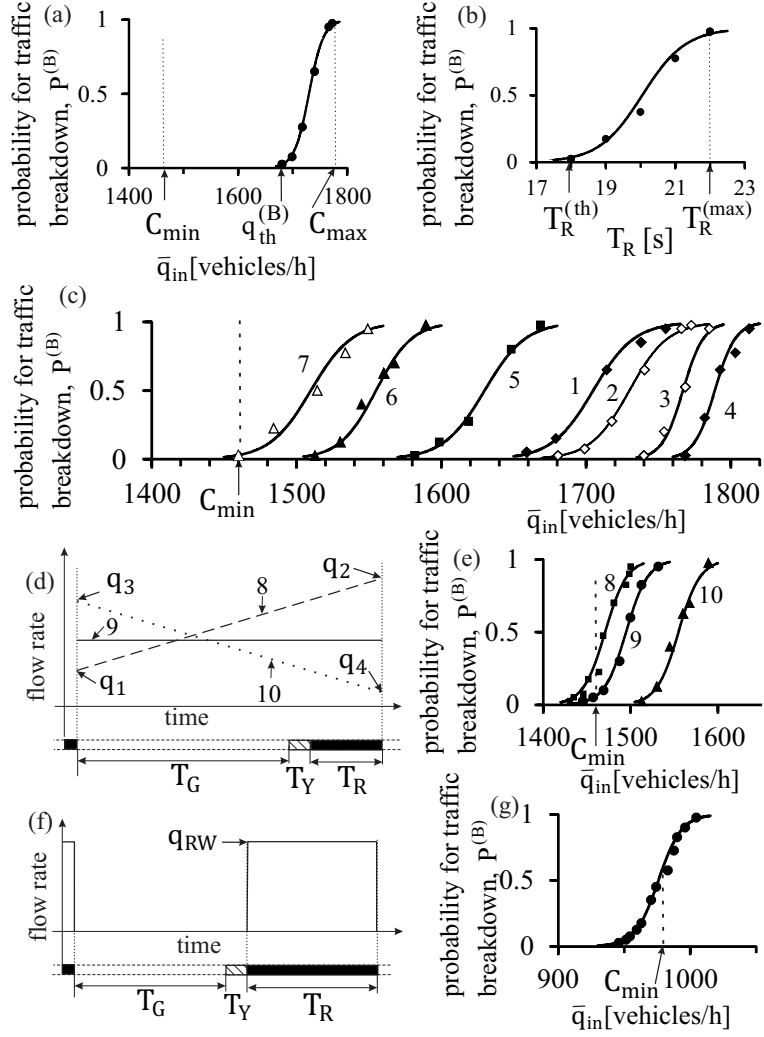}
\caption{Probability of   breakdown at light signal: (a, b) Breakdown probability
   $P^{(\rm B)}$ for $\Delta T^{\rm (ideal)}_{\rm b}=$ 3 s as a function of average
  flow rate  $\bar q_{\rm in}$ (a) at $T_{\rm R}=$ 20 s
and as a function of $T_{\rm R}$ at $q_{\rm GW}=$ 2316 vehicles/h (b).
(c) Comparison of  functions $P^{\rm (B)}(\bar q_{\rm in})$  for
$\Delta T^{\rm (ideal)}_{\rm b}=$ 0 (curve 1), 3 s (curve 2), 6 s (curve 3),   8 s (curve 4
at $x_{\rm LS}=$     1375 m~\cite{Long}),  10 s (curve 5)
as well as turning-in traffic (Sec.~\ref{GW_R_S}) associated with
 the change in $q_{\rm turn}$ (curves 6 and 7 related to $q_{\rm GW}$ given in
Figs.~\ref{Hom_GW_turn_2000_400_3} and~\ref{Hom_GW_turn_1714_1200_3}, respectively).
(d, e) Functions $P^{\rm (B)}(\bar q_{\rm in})$ (e) for different time-dependences
 $q_{\rm in}(t)$ (d) shown by the same numbers   in (d) and (e); in (d), $q_{1}=0.35 q_{2}$, $q_{4}=0.35 q_{3}$.
(f, g) $\lq\lq$Red wave" of duration $T_{\rm R}$ (f)
 and related function $P^{\rm (B)}(\bar q_{\rm in})$ (g).
   $T_{\rm R}=$ 10  (a, c, e) and 
    52 s (g). Other model parameters are the same as those in Fig.~\ref{GW_3_2316_MSP}.
 $C_{\rm min}=$ 
1461 in (a, c, e) and 979 vehicles/h in (g) (see Sec.~\ref{Capacity_S}).   
  \label{Probability_FS} } 
\end{center}
\end{figure}

  In each of the scenarios    discussed above, traffic breakdown    occurs at the light signal  during the time interval  
  $T_{\rm ob}=$ 60 min with some probability $P^{\rm (B)}<1$ {\it only}~\cite{Probability}. This means that
  in some of the different numerical realizations (runs) made the breakdown does occur, however, 
  in other realizations the breakdown does not occur~\cite{Realization}. We have found the following general results:
  
 1. At given model parameters,   $P^{\rm (B)}(\bar q_{\rm in})$ is an increasing
 function on the  flow rate  
 $\bar q_{\rm in}=\vartheta^{-1}\int^{\rm \vartheta}_{0}{q_{\rm in}(t)}dt$ (Figs.~\ref{Probability_FS} (a, c, e))).
  At a given    $q_{\rm in}(t)$,   $P^{\rm (B)}(T_{\rm R})$ is also an increasing function of
    $T_{\rm R}$ (Fig.~\ref{Probability_FS} (b))).
  Functions  
  $P^{\rm (B)}(\bar q_{\rm in})$  and $P^{\rm (B)}(T_{\rm R})$
   can be  fitted with  
  \begin{eqnarray}
  \label{P_q_1}
  P^{(\rm B)}(\bar q_{\rm in})=[1+\exp[\beta (q_{\rm p}-\bar q_{\rm in})] ]^{-1},   \\
  P^{(\rm B)}(T_{\rm R})=[1+\exp[\beta_{\rm R} (T_{\rm p}-T_{\rm R})] ]^{-1},  
  \label{P_q_2}
  \end{eqnarray}
where   $\beta$ and $q_{\rm p}$   depend on   characteristics of   function $q_{\rm in}(t)$ 
and light signal parameters; $\beta_{\rm R}$ and $T_{\rm p}$   depend on   $\bar q_{\rm in}$
and      $\vartheta$.

 2. When  for a green wave  (Fig.~\ref{GW_model}),
 $\Delta T^{\rm (ideal)}_{\rm b}$  increases from  0 to 8 s, 
 function $P^{\rm (B)}(\bar q_{\rm in})$ moves to larger  values $\bar q_{\rm in}$
 (curves 1--4 in Fig.~\ref{Probability_FS} (c)):
 The longer $\Delta T^{\rm (ideal)}_{\rm b}$, the smaller the speed disturbance at the begin of the green wave. 
 This results in   shorter
 vehicle delays. 
 However, this effect  has a limit:  
 At a given $T_{\rm GW}$, the increase in
 $\Delta T^{\rm (ideal)}_{\rm b}$  leads to a decrease in $\Delta T^{\rm (ideal)}_{\rm e}$. Therefore, already a relatively short vehicle delays
can cause the stop of a vehicle(s)  from  the end of the green wave  at the light signal. 
    This explains why  the whole function $P^{\rm (B)}(\bar q_{\rm in})$ begins to 
 move back to smaller flow rates $\bar q_{\rm in}$ (curve 5 in Fig.~\ref{Probability_FS} (c)).
 This shift of $P^{(\rm B)}(\bar q_{\rm in})$ to the smaller $\bar q_{\rm in}$ increases when
  turning-in traffic occurs: The smaller the relation $q_{\rm GW}/q_{\rm turn}$, the larger the shift (curves 6 and 7 in Fig.~\ref{Probability_FS} (c)).

 3. In general, the shift of $P^{(\rm B)}(\bar q_{\rm in})$ to the smaller $\bar q_{\rm in}$ is the more,   
 the longer the queue  build during the previous red phase. This effect is shown
 in   Fig.~\ref{Probability_FS} (d, e) for three different periodic functions $q_{\rm in}(t)=q_{\rm in}(t+\vartheta)$ associated with
 an increase in $q_{\rm in}(t)$ over time   (curve 8), time-independent
 flow rate (curve 9), and a decrease in  $q_{\rm in}(t)$ (curve 10).   
We have found that  the larger  the relation
 $q^{\rm (green)}_{\rm in}/q^{\rm (red)}_{\rm in}$, the more the shift of the function $P^{(\rm B)}(\bar q_{\rm in})$
 to larger   $\bar q_{\rm in}$ (Fig.~\ref{Probability_FS} (e)), where
$q^{\rm (red)}_{\rm in}=T_{\rm R}^{-1}\int^{\vartheta}_{\vartheta - T_{\rm R}}{q_{\rm in}(t)}dt$ and
  $q^{\rm (green)}_{\rm in}=(\vartheta - T_{\rm R})^{-1}\int^{\vartheta - T_{\rm R}}_{0}{q_{\rm in}(t)}dt$.

\section{Infinite number of   capacities of light signal \label{Capacity_S}}

 Traffic  capacity of the light signal  $C$ is defined as   the average flow rate downstream of the light signal  
 $\bar q_{\rm LS}$  
at which traffic   breakdown can occur.

For each set of  a given time-dependence of arrival flow rate  $q_{\rm in}(t)$
   and    light signal     parameters  
  there are the infinite number of such capacities, which are
within the   range
  \begin{equation}
  C_{\rm min}  \leq C \leq C_{\rm max},
  \label{range_c_f}
  \end{equation}
   where $C_{\rm min}$ is  
   the classical capacity, i.e.,  $C_{\rm min}=C_{\rm cl}$ (\ref{cl_cap}), 
   which we call the minimum capacity,
   and $C_{\rm max}$ is the maximum capacity associated with
   the occurrence of spontaneous breakdown at the light signal.

 We define      spontaneous breakdown as a random time-delayed transition from 
  under-  to over-saturated traffic.   
   All    examples  presented above are related to spontaneous
    breakdown. For each given  time-dependence 
   $q_{\rm in}(t)$ 
  and given light signal parameters, spontaneous breakdown occurs 
  with probability
  $P^{\rm (B)}(\bar q_{\rm in})>0$  during the time interval $T_{\rm ob}$ within a range of  $\bar q_{\rm in}$ 
  (Fig.~\ref{Probability_FS} (a))
  \begin{equation}
  q^{\rm (B)}_{\rm th} \leq \bar q_{\rm in} \leq C_{\rm max},
  \label{range_p_f}
  \end{equation}
    where $q^{\rm (B)}_{\rm th}$ is a threshold flow rate for spontaneous breakdown:
    at $\bar q_{\rm in} < q^{\rm (B)}_{\rm th}$ breakdown probability $P^{\rm (B)}(\bar q_{\rm in})=0$. 
    The   maximum capacity $C_{\rm max}$ is defined as the average
   flow rate $\bar q_{\rm in}$ at which  breakdown probability $P^{\rm (B)}$
   reach 1 during the time interval $T_{\rm ob}$ : $P^{\rm (B)}\mid_{\bar q_{\rm in} =C_{\rm max}} =1$. 
    The sense of   maximum capacity is as follows: Under conditions $q^{\rm (B)}_{\rm th} \leq \bar q_{\rm in} < C_{\rm max}$,
   spontaneous breakdown can occur during the time interval
   $T_{\rm ob}$, however, with probability $P^{\rm (B)}<1$. This means that
   in some of realizations~\cite{Realization} no breakdown occurs; therefore, the {\it maximum} capacity is not still reached. Contrarily,
when      $ \bar q_{\rm in} =C_{\rm max}$, then during  the time interval
   $T_{\rm ob}$ spontaneous breakdown does definitely occur.

     \begin{figure}
\begin{center}
\includegraphics*[width=12 cm]{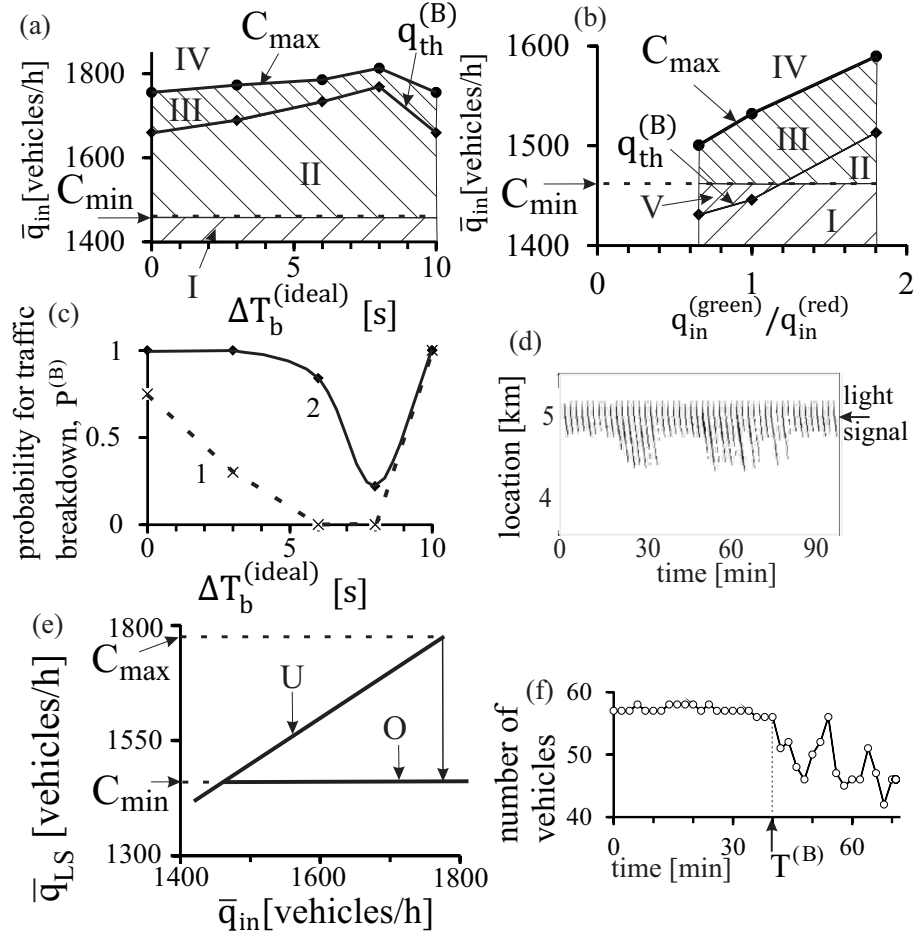}
\caption{Characteristics of  the infinite number of capacities at light signal: (a, b) 
Diagrams of breakdown for green wave   
associated with  functions $P^{(\rm B)}(\bar q_{\rm in})$ shown by curves 1--5 in Fig.~\ref{Probability_FS} (c) and curves 8--10 in
 Fig.~\ref{Probability_FS} (e), respectively, for (a) and (b).
(c) Probability of spontaneous breakdown as function of $\Delta T^{\rm (ideal)}_{\rm b}$ for two given flow rates
$\bar q_{\rm in}=$ 1720 (dashed curve 1) and 1780 vehicles/h (solid curve 2) associated with (a).
(d)   Speed data in the space-time plane of
  for dissolving over-saturated traffic occurring within region V in diagram
(b) for function $q_{\rm in}(t)$ shown by curve 8 in Fig.~\ref{Probability_FS} (d) for $\bar q_{\rm in}-C_{\rm min}=
- 15$ vehicles/h, $q_{1}=$ 746 and $q_{2}=$ 2130 vehicles/h.
(e, f) Flow--flow characteristics $\bar q_{\rm LS}(\bar q_{\rm in})$ (e)
and dependence of number of vehicles passing the light signal (f) for green waves with  $\Delta T^{\rm (ideal)}_{\rm b}=$ 3 s
related to Fig.~\ref{GW_3_2316_MSP}.    
  \label{Capacity} } 
\end{center}
\end{figure}

   \section{Induced traffic breakdown and double Z-characteristic  \label{Induced_S}} 

The minimum capacity $C_{\rm min}$ can be considerably smaller than 
 $ q^{\rm (B)}_{\rm th}$ (Fig.~\ref{Capacity} (a)).  However,  under condition
$\bar q_{\rm in} < q^{\rm (B)}_{\rm th}$ probability of
 spontaneous breakdown is equal to zero. Nevertheless, in accordance with (\ref{range_c_f})
within the flow rate range 
    \begin{equation}
  C_{\rm min} \leq \bar q_{\rm in} < q^{\rm (B)}_{\rm th}
  \label{ind_f}
  \end{equation}
under-saturated traffic is in a metastable state with respect to the transition to over-saturated traffic.
Therefore, in this flow rate range
the breakdown   can be induced by external time-limited disturbances in  under-saturated traffic. 
Induced breakdown can occur even if a disturbance appears
during {\it only one} of the light signal cycles.
Examples of such disturbances are a random deceleration of one of the vehicles within a green wave or
a queue caused by turning-in traffic.

       \begin{figure}
\begin{center}
\includegraphics*[width=11 cm]{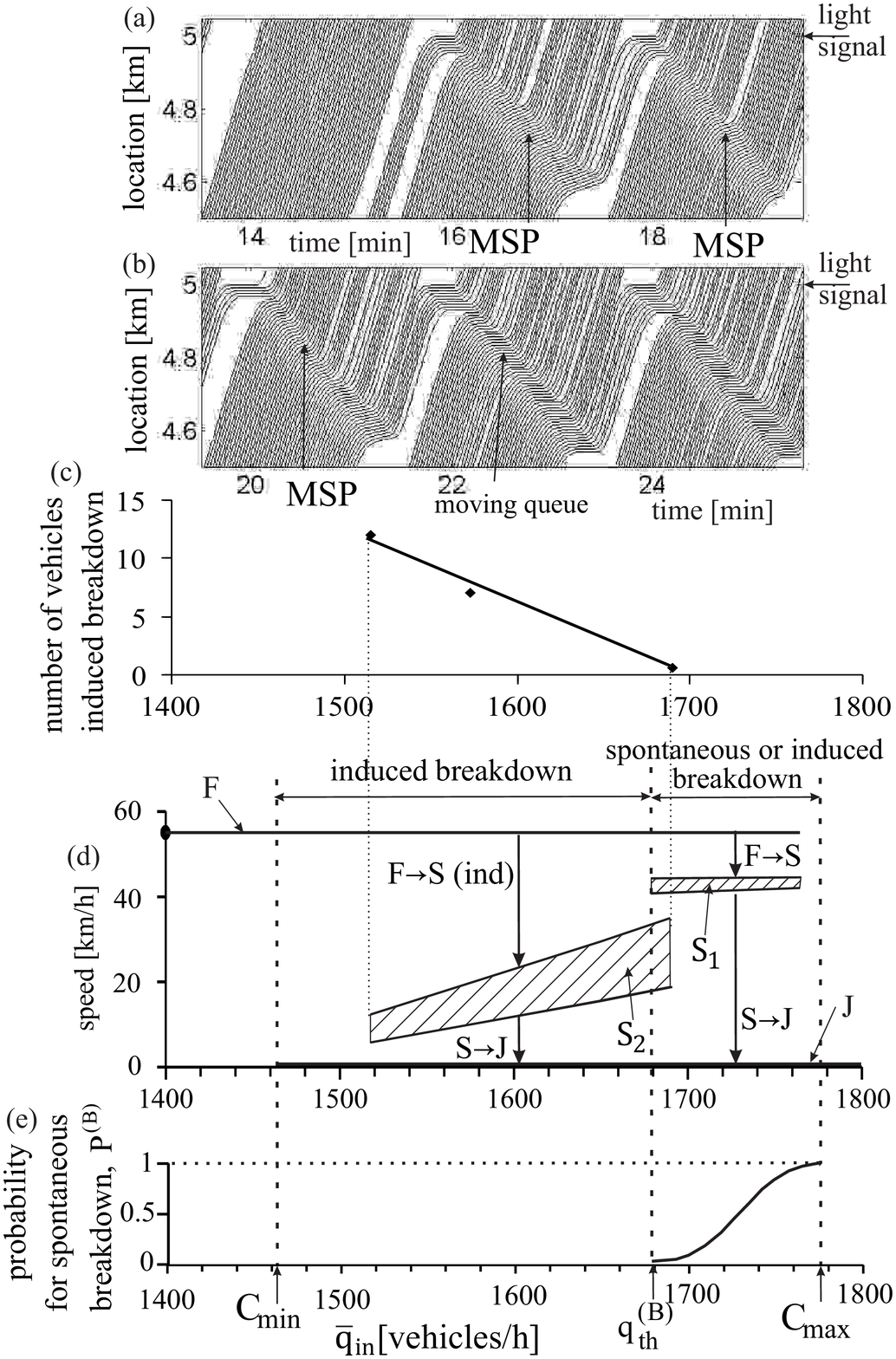}
\caption{Induced traffic breakdown at light signal:  
(a, b) Vehicle trajectories for $q_{\rm GW}=$ 2057 vehicles/h
and $M_{\rm turn}=$ 7 vehicles.  (c) Number of vehicles $M_{\rm turn}$ 
  inducing    breakdown as function of  $\bar q_{\rm in}$.  
  $(q^{\rm (B)}_{\rm th}, \ C_{\rm max})=$ (1680, 1772)    vehicles/h. (d)
  Double Z-characteristic.  
  (e) Breakdown probability taken from Fig.~\ref{Probability_FS} (a). In (d),
regions  $S_{1}$ and $S_{2}$   are related to synchronized flow within MSPs.
  Other parameters the same as those in Fig.~\ref{GW_3_2316_MSP}.
  \label{Induced} } 
\end{center}
\end{figure}

 For a green wave  (Fig.~\ref{GW_model}), we choose the flow rate $\bar q_{\rm in}=q_{\rm GW}(T_{\rm GW}/\theta)$ satifying
 condition (\ref{ind_f}) (Fig.~\ref{Induced}). Then $P^{\rm (B)}=0$, i.e., no  green wave breakdown can occur spontaneously.
 Now, during the red phase of {\it the only  one cycle}
  $M_{\rm turn}$ vehicles appear due to turning-in traffic ($M_{\rm turn}=$7 
  at  $t\approx $ 15.7 min in Fig.~\ref{Induced} (a)). The vehicles
   are build a queue during the   red phase. The  discharge
  of the queue  at the following green phase
  causes a speed disturbance  occurring  within the associated green wave.
   We have found the following   phenomena:
   (i) For each given $\bar q_{\rm in}$ that satisfies (\ref{ind_f})
   there is a   value of $M_{\rm turn}$
   at which the
   disturbance induces breakdown with some probability during the time
   interval $T_{\rm ob}$~\cite{Prob_ind}. The smaller $\bar q_{\rm in}$,
   the larger $M_{\rm turn}$ is required for the breakdown (Fig.~\ref{Induced} (c)).
   (ii) The speed disturbance causes the emergence of an MSP within the green wave.
   (iii) The subsequent development of the MSP, which is qualitatively the same as that for spontaneous breakdown
   (Sec.~\ref{Br_S}), leads to over-saturated traffic. 
    
  A sequence of F$\rightarrow$S$\rightarrow$J transitions
 at the light signal (Secs.~\ref{GW_S} and~\ref{GW_R_S})  can be presented in the speed--flow-rate plane  
  by a double Z-characteristic for phase transition
 in traffic at the light signal (Fig.~\ref{Induced} (d)), which exhibits the following characteristics: (i) 
  An F$\rightarrow$S transition with MSP emergence shown by arrow $\lq\lq$F$\rightarrow$S".
(ii) An S$\rightarrow$J transition
shown by arrow $\lq\lq$S$\rightarrow$J"~\cite{Ind_F_J}.
  (iii) Under condition (\ref{range_p_f}), spontaneous F$\rightarrow$S$\rightarrow$J transitions
  can occur with probability $P^{\rm (B)}(\bar q_{\rm in})>0$  during the time interval $T_{\rm ob}$
  (Fig.~\ref{Induced} (e)).
  (iv) Under condition (\ref{ind_f}),  an F$\rightarrow$S transition 
  (labeled by $\lq\lq$F$\rightarrow$S (ind)")
  can be induced. (v) Under condition (\ref{range_p_f}), regions of induced and spontaneous breakdowns are partially
  overlapping each other: 
the breakdown can be induced before spontaneous breakdown  occurs.

  \section{Diagram of traffic breakdown at light signal  \label{Diagram_S}}

  A diagram of the  breakdown  
   presents  regions of the  flow rate $\bar q_{\rm in}$,  
  within which the breakdown  can occur, in dependence of
     light signal parameters  or/and
  parameters of the time-function $q_{\rm in}(t)$ (Fig.~\ref{Capacity} (a, b)).
 Regions I--V in the diagrams   (Fig.~\ref{Capacity} (a, b)) have
  the following meaning: I is related to stable under-saturated traffic,
  II -- metastable under-saturated traffic, III -- metastable under-saturated traffic in which spontaneous
  breakdown can occur, IV -- unstable under-saturated traffic, and in region V  {\it dissolving} over-saturated traffic can occur.
  In   dissolving over-saturated traffic, random emergence and subsequent dissolution of
 a growing queue   at the light signal follows each other randomly (Fig.~\ref{Capacity} (d)).

   The maximum capacity $C_{\rm max}$, which   
  determines top diagram boundary, 
  can exhibit
  a maximum as a function of light signal parameters (Fig.~\ref{Capacity} (a)). An analysis of this diagram     shows that
  there is a  minimum of  
 breakdown  probability  $P^{\rm (B)}$ 
   as a function of $ \Delta T^{\rm (ideal)}_{\rm b}$ 
  at given other parameters (Fig.~\ref{Capacity} (c)).

  \subsection{$\lq\lq$Red" wave: Transition to classical definition of capacity at light signal} 
  
  The smaller the relation $q^{\rm (green)}_{\rm in}/q^{\rm (red)}_{\rm in}$, the smaller the difference
  $C_{\rm max}-C_{\rm min}$ (Fig.~\ref{Capacity} (b)). 
  In the limit case $q^{\rm (green)}_{\rm in}/q^{\rm (red)}_{\rm in} =0$, which we call $\lq\lq$red" wave because
  all vehicles arrive the light signal during the red phase
  only  (Fig.~\ref{Probability_FS} (f)), the difference  
  $C_{\rm max}-C_{\rm min}$ becomes very small. Therefore,
   the transition from under- to over-saturated traffic occurs on average at
 $\bar q_{\rm in}=C_{\rm min}$  as stated in the classical traffic flow theories~\cite{Webster,Gartner,Rakha,Fluc_Cap}.
 Thus only for the red wave one can determine  traffic capacity at the light signal
  based on the classical capacity definition   (Sec.~\ref{Int}).  
This emphasizes that and why in all realistic cases in which  during the green and yellow phases
$q_{\rm in}(t)\neq 0$  there are the infinite number of capacities at the light signal
within the capacity range (\ref{range_c_f}).

  \subsection{Flow--flow characteristic of green wave breakdown}
  
  A flow--flow characteristic explains the evolution of green wave in the flow--flow plane with
  coordinates $(\bar q_{\rm LS}, \bar q_{\rm in})$, where
 $\bar q_{\rm LS}={\vartheta}^{-1}\int^{\vartheta}_{0}{q_{\rm LS}(t)}dt$ 
 is average  rate in the light signal outflow, i.e., downstream of the light signal (Figs.~\ref{GW_model} 
 and~\ref{Capacity} (e)):
If  $\bar q_{\rm in}$     (Sec.~\ref{GW_M_S})
increases beginning from   small values,   $\bar q_{\rm LS}= \bar q_{\rm in}$
(branch $U$ for under-saturated traffic in Fig.~\ref{Capacity} (e)).
Under-saturated traffic associated with green wave can
exist   even when $\bar q_{\rm in}>C_{\rm min}$. However, at
$\bar q_{\rm in}=C_{\rm max}$ during the time interval $T_{\rm ob}$ with  probability 
     $P^{\rm (B)}=1$  the green wave breakdown does occur: The green wave destroys resulting in the decrease in the outflow rate  
 from $\bar q_{\rm LS}=C_{\rm max}$    to $\bar q_{\rm LS}=C_{\rm min}$ (arrow from
      branch $U$ to branch $O$ for over-saturated traffic) caused by the breakdown.  
     Before   green wave breakdown occurs, the number of vehicles passing the light signal  
      is almost time-independent ($t<T^{\rm (B)}$ in Fig.~\ref{Capacity} (f)); after the breakdown  
     it exhibits a very complex time behavior  ($t>T^{\rm (B)}$).

     After over-saturated traffic  has  occurred, 
     the presented theory shows the well-known result
     of the classical theory~\cite{Webster,Gartner,Rakha}: When over-saturated traffic exists at the light signal
     (branch $O$ in Fig.~\ref{Capacity} (e)),  a large decrease in
     $\bar q_{\rm in}$ to $\bar q_{\rm in}<C_{\rm min}$ 
     is needed for the return transition to under-saturated traffic.

  \section{Breakdown of green wave at sequence of light signals \label{Many_LS_S}}

\begin{figure}
\begin{center}
\includegraphics*[width=10 cm]{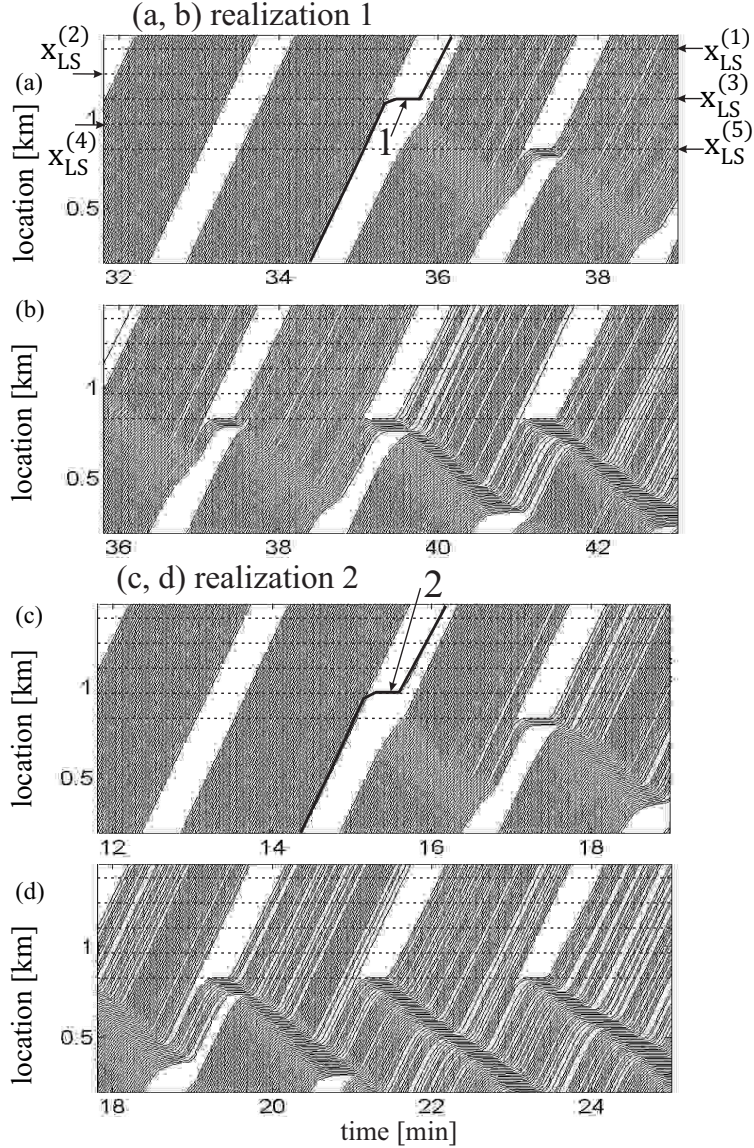}
\caption{Green wave breakdown at five light signals at $\vartheta=$ 120 s, $T_{\rm R}=$ 20 s,  $T_{\rm GW}=$90
s, and $\Delta T^{\rm (ideal)}_{\rm b}= $ 8 s:   Vehicle trajectories for realization 1 (a, b) and 2 (c, d). 
$\Delta x_{\rm LS}=$  137.5 m, $x^{(1)}_{\rm LS}-x_{\rm b}=$ 1375 m~\cite{Long}.
 $q_{\rm GW}= 2382$ vehicles/h.
  \label{GW_3_2292_five3} } 
\end{center}
\end{figure}

\begin{figure}
\begin{center}
\includegraphics*[width=10 cm]{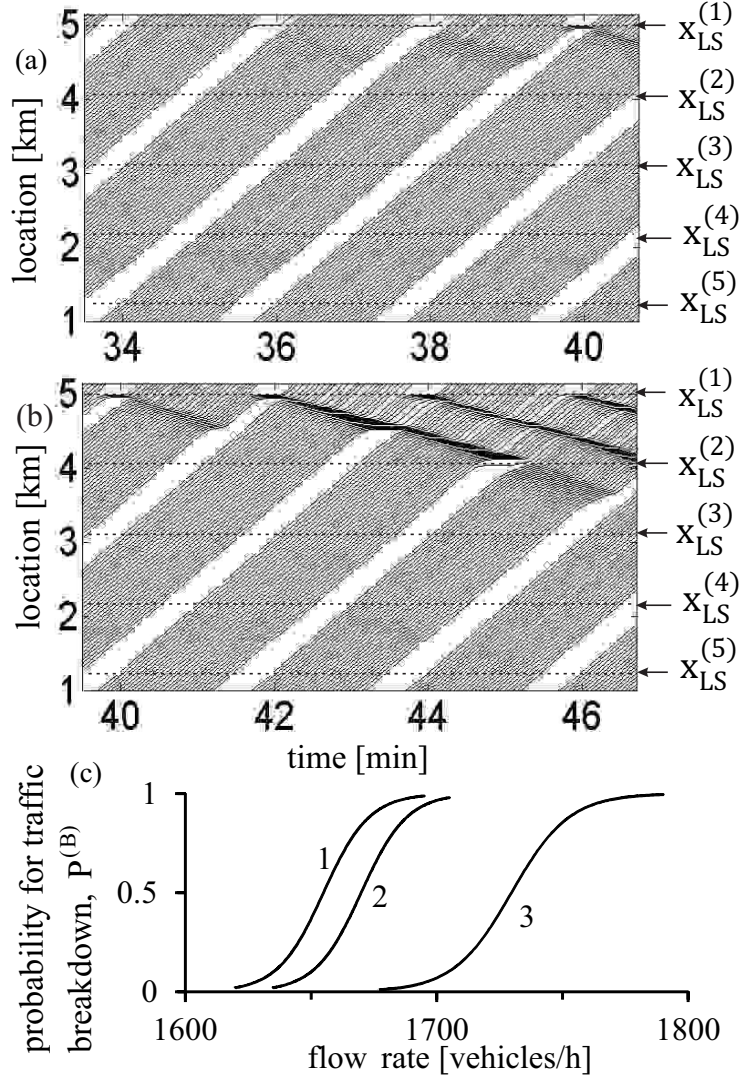}
\caption{Green wave breakdown at five light signals at $\vartheta=$ 120 s, $T_{\rm R}=$ 20 s,  $T_{\rm GW}=$90
s, and $\Delta T^{\rm (ideal)}_{\rm b}= $ 3 s: (a, b) Vehicle trajectories; 
$\Delta x_{\rm LS}=$   962.5 m~\cite{Long}.
(c) Functions
$P^{\rm (B)}(\bar q_{\rm in})$ of
probability   that green wave breakdown occurs at {\it  one of the   light signals} during the time interval $T_{\rm ob}$ 
(curves 1 and 2 for $\Delta x_{\rm LS}=$ 687.5  and
 137.5   m, respectively) and their comparison with   $P^{\rm (B)}(\bar q_{\rm in})$ for isolated
light signal (curve 3 taken from Fig.~\ref{Probability_FS} (a)). In (a, b), 
within time interval $0<t<34$ min for different green waves we find the following ranges of random changes in  time gaps
$(\Delta T^{\rm (p)}_{\rm b}, \ \Delta T^{\rm (p)}_{\rm e})=$ (5.32--5.93, 0.05--3.41), (5.11--5.71, 0.28--3.9),
(4.8--5.4, 0.81--4.99), (4.4--4.8, 0.87--5.3), and (3.79--4.43, 2.77--6.78) s
for ${\rm p=1, 2,..., 5}$, respectively; the mean values of these time gaps are (5.61, 1.99),
(5.3, 2.33), (5.01, 2.95), (4.56, 3.87), and (3.95, 5.13) s for ${\rm p=1, 2,..., 5}$, respectively. 
 $q_{\rm GW}= 2252$ vehicles/h.
  \label{GW_3_2292_five2} } 
\end{center}
\end{figure}

\begin{figure}
\begin{center}
\includegraphics*[width=9 cm]{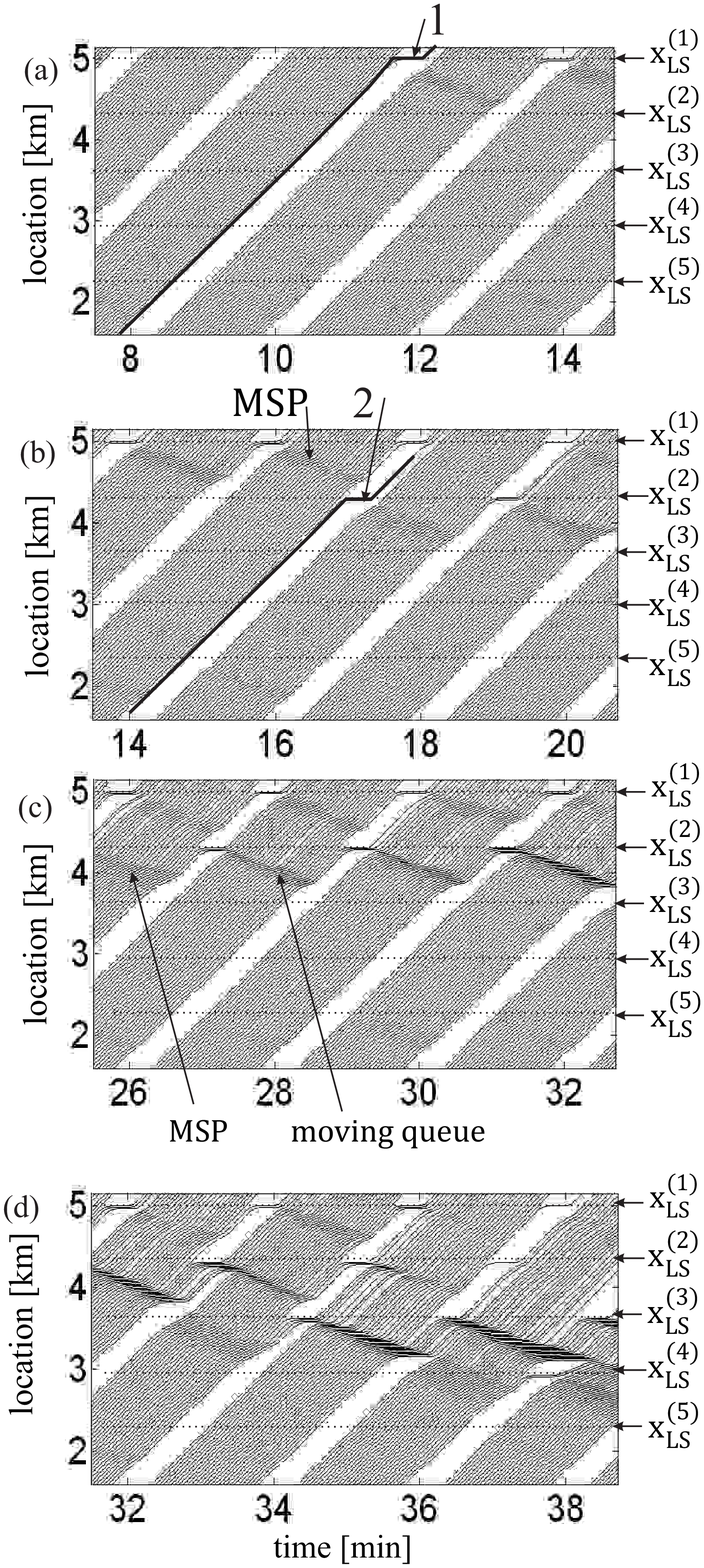}
\caption{Green wave breakdown at five light signals at $\vartheta=$ 120 s, $T_{\rm R}=$ 20 s,  $T_{\rm GW}=$90
s, and $\Delta T^{\rm (ideal)}_{\rm b}= $ 3 s:   Vehicle trajectories. 
$\Delta x_{\rm LS}=$  687.5 m~\cite{Long}.
 $q_{\rm GW}= 2252$ vehicles/h.
  \label{GW_3_2292_five} } 
\end{center}
\end{figure}

  We consider  green wave propagation 
 through a sequence of the light   signals 
 at equidistant locations $x=x^{\rm (p)}_{\rm LS}$ with 
a  distance between them $\Delta x_{\rm LS}$ and a time shift of the green phase beginning
  $\Delta T_{\rm G}=\Delta x_{\rm LS}/v_{\rm free}$,
 where $p=1, 2, 3,..., P$; $P>1$ is the number of the light signals.  
 We have revealed the following   
results   (Figs.~\ref{GW_3_2292_two}--\ref{GW_3_2292_five}):
(i) In a neighborhood of {\it each of} the light signals
 an MSP can  
occur. Because vehicles   move through  MSPs occurring at different
 light signals,  the
 mean values of random  time gaps $\Delta T_{\rm b}$ and $\Delta T_{\rm e}$  
depend on the light signal location (see caption to Fig.~\ref{GW_3_2292_five2}).   
(ii) Stages of the green wave breakdown  are
 the same as those found for the isolated light signal (Sec.~\ref{Br_S}).
(iii) However, there is a stochastic {\it dynamic  competition}
  in the   development
  of the green wave breakdown {\it between}   light signals. In particular,
  it  turns often out that although a queue appears firstly  at one of the light signals (stage (ii) of Sec.~\ref{Br_S}), 
  the breakdown is realized (stage (iv) of Sec.~\ref{Br_S}) at another one.  
Characteristic features of this   dynamics depend 
on values $\Delta T^{\rm (ideal)}_{\rm b}$ and
$\Delta x_{\rm LS}$    (Figs.~\ref{GW_3_2292_two}--\ref{GW_3_2292_five}):

     \begin{figure}
\begin{center}
\includegraphics*[width=10 cm]{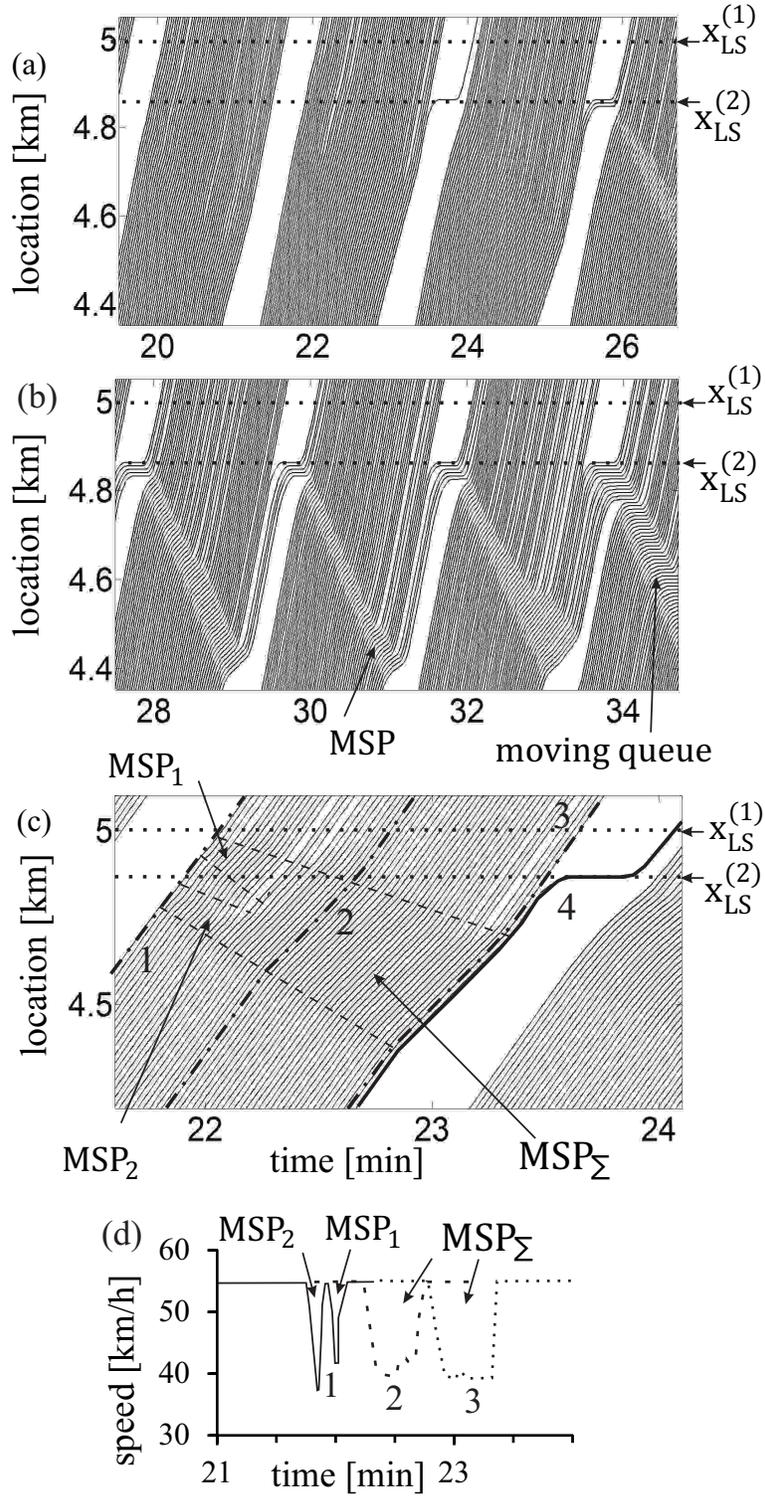}
\caption{Green wave breakdown at two light signals at $\vartheta=$ 120 s, $T_{\rm R}=$ 20 s,  $T_{\rm GW}=90$
s, and $\Delta T^{\rm (ideal)}_{\rm b}= $ 3 s: (a--c) Vehicle trajectories. (d) Microscopic speeds of vehicles
moving thorough MSPs labeled in (c).
$\Delta x_{\rm LS}=$ 137.5 m~\cite{Long}.
 $q_{\rm GW}= 2292$ vehicles/h.
  \label{GW_3_2292_two} } 
\end{center}
\end{figure}

1. If  $\Delta x_{\rm LS}$ is considerably shorter than the distance that an MSP propagates
 to the green wave end    (about 400 m in
Fig.~\ref{GW_3_2316_MSP}), then with the largest probability the breakdown 
occurs at one of the upstream light signals. In  Fig.~\ref{GW_3_2292_two}, vehicles within the green wave
should propagate through an MSP occurring at the upstream light signal that they approach
($MSP_{2}$ in Fig.~\ref{GW_3_2292_two} (c, d)) {\it and} through another MSP occurring at the subsequent downstream light signal
($MSP_{1}$). Both MSPs merge
within the green wave into one MSP with a larger width ($MSP_{\sum}$ in Fig.~\ref{GW_3_2292_two} (c, d)). 
This increases vehicle delays within the MSPs resulting in the breakdown  at the upstream light signal.

2. Under condition (\ref{dist_non})  no MSPs appear initially at the beginning of green waves (Sec.~\ref{Green_along_S}).
 Although at a short value $\Delta x_{\rm LS}$ with the largest probability the breakdown 
occurs also at one of the upstream light signals (Fig.~\ref{GW_3_2292_five3} (b, d)), 
  in different realizations~\cite{Realization} the stop of a vehicle(s)
initiating the breakdown process is observed  at different light signals (trajectories 1 and 2
 in Fig.~\ref{GW_3_2292_five3} (a) and (c), respectively).

3. If  
$\Delta x_{\rm LS}$ is   long enough, then with the largest probability the breakdown 
occurs at one of the  downstream light signals (Fig.~\ref{GW_3_2292_five2} (a, b)): Approaching the furthest downstream light signal  vehicles   exhibit
the   longest mean time delay
caused by     MSPs  
within the green wave.

4. When $\Delta x_{\rm LS}$ is 
comparable with
the distance that the MSP propagates
  to the green wave end, the stop of  last vehicles of the green wave
  (stage (ii) of Sec.~\ref{Br_S}) can occur    at several
  neighborhood light signals.  
   In example shown in Fig.~\ref{GW_3_2292_five},  
 the stop of the last vehicle of a green wave  occurs  at the downstream  
light signal (trajectories  1 in Fig.~\ref{GW_3_2292_five} (a)).  This results in  MSPs 
  emerging  at the downstream light signal.  One of the MSPs
  (MSP marked in Fig.~\ref{GW_3_2292_five} (b)) can cause
the vehicle stop  at the neighborhood upstream light signal
 (trajectory  2 in Fig.~\ref{GW_3_2292_five} (b)). In turn, this  vehicle stop  decreases   the  number of vehicles within the green wave
  approaching the downstream light signal. This results in
the interruption of   the breakdown process
at this light signal: The breakdown process that has started at the downstream light signal leads to the breakdown
at one or a few of the upstream light signals (Fig.~\ref{GW_3_2292_five} (c, d)).

   \section{Applications of breakdown minimization (BM) principle
  for optimization of green wave in a city \label{BM_S}}

 For a traffic network with  $N$   bottlenecks  the
  BM principle  is as follows~\cite{BM,Kerner2011_TEC,Kerner2011_ITS}: 
  The network optimum is reached, when dynamic traffic optimization and/or control are performed in the network in such a way that 
  the probability for spontaneous occurrence of traffic breakdown 
  in at least one of the network bottlenecks during a given observation time  $T_{\rm ob}$
  reaches the minimum possible value. The BM principle 
    is equivalent to the maximization of 
  the probability that traffic breakdown occurs at {\it none} of the network bottlenecks.

Assuming that    traffic breakdown at different bottlenecks in the network is independent each other,
the probability 
for spontaneous occurrence of
 traffic breakdown in at least   one of the network bottlenecks during  the time interval  $T_{\rm ob}$  can be written as:
\begin{eqnarray}
P_{\rm net}=  1- \prod^{\rm N}_{k=1} (1-P^{\rm (B,{\it k})}).
\label{P_net_BM}
\end{eqnarray}
In accordance with the BM principle,  the network optimum is reached at~\cite{BM,Kerner2011_TEC,Kerner2011_ITS}
  \begin{equation}
\min \limits_{q_{1},q_{2},...,q_{\rm M},\zeta_{1},\zeta_{2},...,\zeta_{\rm M},\alpha_{1},\alpha_{2},...,\alpha_{W}} \{P_{\rm net}(q_{1},q_{2},...,q_{\rm M},\zeta_{1},\zeta_{2},...,\zeta_{\rm M},\alpha_{1},\alpha_{2},...,\alpha_{W})\},
\label{P_net_BM_1}
\end{equation}
where 
 $M$ is the number of network links for which inflow rates can be adjusted,
$q_{m}$ is the link inflow rate
for a link with index $m$; 
$\zeta_{m}$ is a matrix
of percentages of vehicles with different vehicle (and/or driver) characteristics 
that influence on the breakdown probability at a bottleneck;   
the matrix $\zeta_{m}$ takes into account that dynamic assignment
is possible individually for each of the vehicles~\cite{zeta}; $m=1,2,..., M$, where $M>1$;
$k= 1,2,..., N$ is bottleneck 
index, 
$N> 1$;
$P^{\rm (B,{\it k})}$ is probability that during 
the time  $T_{\rm ob}$ traffic breakdown occurs at  bottleneck   $k$;  $\alpha_{w}$  
is the set of control parameters  
for one of these $N$  bottlenecks with index  $w$ ($w=1,2,...,W$),   $W\leq N$~\cite{LS_cite}.
    The BM principle   is  equivalent to
 \begin{eqnarray}
\max \limits_{q_{1},q_{2},...,q_{\rm M},\zeta_{1},\zeta_{2},...,\zeta_{\rm M},\alpha_{1},\alpha_{2},...,\alpha_{W}} \{P_{\rm C, net}(q_{1},q_{2},...,q_{\rm M},\zeta_{1},\zeta_{2},...,\zeta_{\rm M},\alpha_{1},\alpha_{2},...,\alpha_{W})\},
\label{P_net_BM_2}
\end{eqnarray}    
 where
  \begin{equation}
P_{\rm C, net}=   \prod^{\rm N}_{k=1} P^{\rm (B,{\it k})}_{\rm C}
\label{P_net_BM_3}
\end{equation}  
is the probability that 
 during     time interval $T_{\rm ob}$  free flows remain
   in the  network,  i.e.,   that
traffic breakdown occurs at {\it none} of the  bottlenecks,
  \begin{equation}
P^{\rm (B, {\it k})}_{\rm C}=1-P^{\rm (B, {\it k})}.
\label{P_net_BM_4}
\end{equation}
 
 The existence of a minimum of breakdown probability on the time gap $\Delta T^{\rm (ideal)}_{\rm b}$ (Fig.~\ref{Capacity} (c))
 allows us to suggest some simple applications of the BM principle for the green wave optimization.
In a hypothetical case of a sequence of  light signals $k=1, ..., N_{1}$ that
 are at long enough distances each other  
 the green wave optimization at given $\theta$, $T_{\rm R}$, $T_{\rm G}$, and $T_{\rm GW}$
  can be achieved  by a choice of optimal values $\Delta T^{\rm (ideal)}_{\rm b,{\it k}}, k=1, ..., N_{1}$. 
  Indeed, in the case  the BM principle (\ref{P_net_BM_1}) leads to a simple result
  that  the optimum for green wave  is reached, when each of the breakdown probabilities
     $P^{\rm (B,{\it k})}$ for   the associated light signals
$k=1, ..., N_{1}$      reaches minimum as a function of $\Delta T^{\rm (ideal)}_{\rm b,{\it k}}$.

In the case of a complex   dynamic  competition 
    between the light signals $k=1, ..., N_{1}$ (Sec.~\ref{Many_LS_S}),   traffic breakdowns at 
    these different light signals cannot be considered independent events. 
However, these light signals   we can consider 
   a {\it single bottleneck}.  In  the  BM principle
 (\ref{P_net_BM}), (\ref{P_net_BM_1}),
breakdown  probability $P^{\rm (B,{\it s})}$  for this single bottleneck with some index $k=s$ is associated with  
  probability for the   breakdown  occurring
at {\it one} of the light signals during the time interval $T_{\rm ob}$. 
Simulations show that  for a sequence of the light signals  
the function  $P^{\rm (B)}(\bar q_{\rm in})$ satisfies 
  formula (\ref{P_q_1}) and it is usually
shifted to smaller flow rates in comparison with the function  
  $P^{\rm (B)}(\bar q_{\rm in})$ for an isolated light signal (Fig.~\ref{GW_3_2292_five2} (c)).
This application of the BM principle  is possible only when $N_{1}<N$, i.e., when in addition to
the single bottleneck caused by the light signals 
there are also other
bottlenecks in the network.
   
As introduced in~\cite{BM,Kerner2011_TEC,Kerner2011_ITS},  traffic network optimization through the use of the BM principle 
 can be a combination of a {\it global} network optimization with {\it local} control of 
 a   bottleneck consisting of the following   stages:
 
(i) {\it Global network optimization}: The minimization of traffic breakdown probability in the network based on the BM
principle.

(ii) {\it Local bottleneck control}: A spatial limitation of congestion growth when traffic breakdown has nevertheless
occurred at a network bottleneck, with the subsequent congestion dissolution at
the bottleneck, if the dissolution of congestion due to traffic management in a
neighborhood of the bottleneck   is possible. 

(iii) {\it Combination of global network optimization 
with local control of congested bottlenecks}: 
The minimization of traffic breakdown probability with
the BM principle in
 the network part that is not
influenced by congestion  together with local control of congested bottlenecks mentioned in item (ii).

    \begin{figure}
\begin{center}
\includegraphics*[width=8 cm]{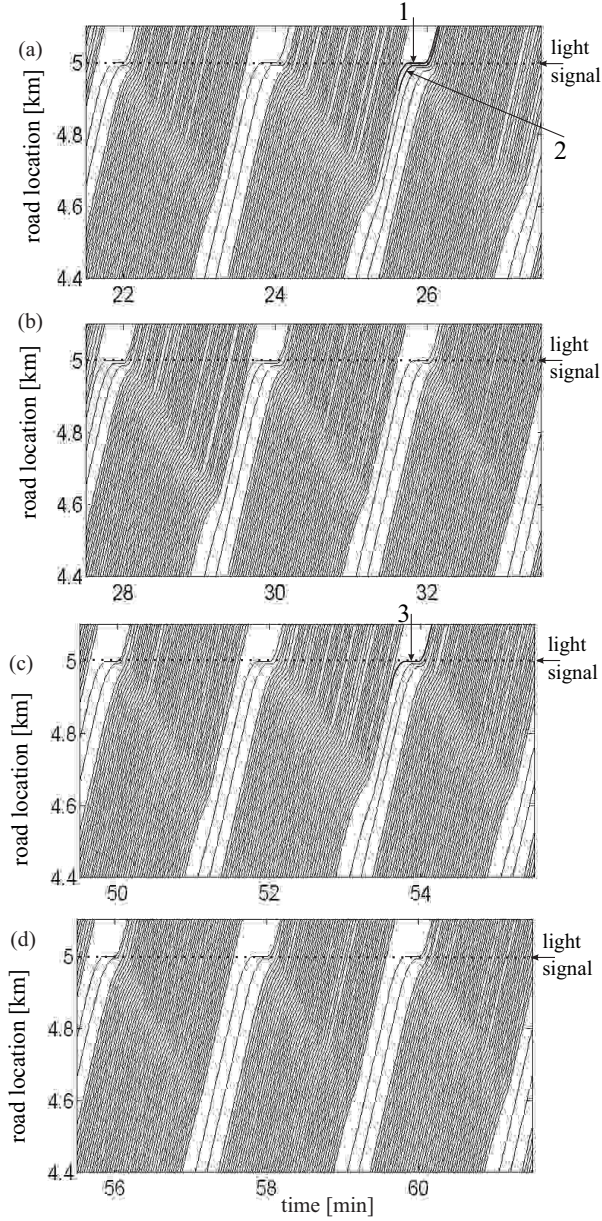}
\caption{Control of green wave breakdown shown in
Fig.~\ref{Hom_GW_turn_2000_400_3}. Vehicle trajectories:  
   (a, b) Interruption of   breakdown process due to   control.  
     After two last vehicles of  green wave
   come to a stop at $t\approx $ 25.8 min (bold   trajectories 1 and 2
  are the same as   1 and 2 in Fig.~\ref{Hom_GW_turn_2000_400_3} (d)), 
    $T_{\rm R}$  is set to be 1 s shorter for the  {\it only}  one light signal  cycle. In (b),
   due to   control  the number of stopped vehicles    decreases at the next cycle (from
  two vehicles to one vehicle) although   
for   subsequent cycles    $T_{\rm R}$    remains to be 20 s. (c, d) Green wave control shown in (a, b) 
repeats each time, when the breakdown process starts   over time (in (c) the last vehicle of the green wave 
comes to a stop at $t\approx$ 53.8 min as shown by bold trajectory 3); as in (a, b), this control  results in    breakdown interruption (d).
\label{Hom_GW_turn_2000_400_3_control} } 
\end{center}
\end{figure}

We see that in the approach of traffic network optimization and control of Ref.~\cite{BM,Kerner2011_TEC,Kerner2011_ITS},
  local bottleneck control   begins only after the process of traffic breakdown has already  started
 at the bottleneck and, therefore, this network bottleneck cannot  further be included in  global network optimization 
 with the BM principle (\ref{P_net_BM_1}).  
 
 Local bottleneck control
 can be very effective for a green wave because  between the  start of the breakdown
 (stage (ii) of Sec.~\ref{Br_S}) and the breakdown instant  (stage (iv) of Sec.~\ref{Br_S}) there can be a long   time interval
 associated with several cycles of the light signal.  Through appropriate control made within this time interval
 the breakdown process can be interrupted as
  shown in Fig.~\ref{Hom_GW_turn_2000_400_3_control}. 
  
  After congestion dissolution or breakdown interruption has been achieved at the bottleneck,
  this bottleneck can   again be included in  global network optimization 
 with the BM principle (\ref{P_net_BM_1}).  
 If rather than  congestion dissolution or breakdown interruption at the bottleneck (Fig.~\ref{Hom_GW_turn_2000_400_3_control})
only the limitation of the congestion growth can be achieved through traffic management in a
neighborhood of a bottleneck, global network optimization with the BM  principle is performed
only for a network part that is not
influenced by congestion: 
  In   (\ref{P_net_BM}), breakdown probabilities
for  only   those network bottlenecks  at which no congestion has occurred
 should be taken into account.

        \section{Discussion \label{Discussion_S}}
     
 \subsection{Comparison of traffic  breakdown at highway bottleneck and  light signal \label{High_Light_S}}

F$\rightarrow$S$\rightarrow$J transitions disclosed above as the reason for
traffic breakdown  at the  light signal  occur because
remaining vehicles stopping at the previous red cycle act
as a disturbance for the next traffic. A large enough on-ramp inflow at an on-ramp bottleneck  
acts  also  as   disturbance for
traffic on the main road   causing  F$\rightarrow$S$\rightarrow$J transitions 
at the highway bottleneck~\cite{Kerner1998B,KernerBook}.
Therefore, questions arise: What new features are induced by the existence of traffic
lights in comparison with those   for the on-ramp bottleneck?
Is the effect of vehicles stopping at the previous red cycle of the light signal
different from
those of in-coming traffic at on-ramp? Responses to these questions are as follows.

1. During the red phase, traffic is interrupted at the light signal resulting in a vehicle queue.
The downstream queue front      is fixed at the light signal: The outflow from this queue
{\it is zero}. In contrast,  during the whole time and independent on     on-ramp inflow 
the outflow from congested traffic at the on-ramp bottleneck
{\it is   not zero}.  This   leads to the following qualitative different
traffic features at the light signal and on-ramp bottleneck:
  
(i) Due to the F$\rightarrow$S transition, a widening SP (WSP) or localized SP (LSP) (SP -- synchronized flow pattern)  
   can occur at the on-ramp bottleneck~\cite{KernerBook}. The downstream front of the WSP or LSP is fixed at the bottleneck.
   Within this front  
   vehicles accelerate from synchronized flow upstream of the on-ramp bottleneck to free flow downstream. The existence of the WSP or LSP is
   possible because  the on-ramp bottleneck does not interrupt traffic flow.
    In contrast, the light signal   interrupts traffic flow
     during the red phase. 
For this reason, neither WSP nor LSP  can  occur     
at the light signal.
  
 (ii) Rather than
   WSP  or  LSP emergence,    the F$\rightarrow$S transition occurring 
 in arrival traffic  during the green phase leads to an  MSP whose
   downstream front   propagates upstream of the light signal.

(iii) During the green phase, the downstream front of the queue  moves upstream
as those for a wide moving jam in highway traffic. Therefore, this moving queue is a synonym of
the wide moving jam.
 However,  
the moving queue occurs at the light signal. In contrast, wide moving jams emerge at some distance from the on-ramp bottleneck
   location at which the F$\rightarrow$S transition has initially occurred~\cite{KernerBook,Heavy}.

2. Traffic breakdown  at the on-ramp bottleneck is an F$\rightarrow$S transition:
After the F$\rightarrow$S transition
has occurred, synchronized flow (one of the phases of congested traffic) remains at the bottleneck.
In contrast, traffic breakdown at the light signal is associated with F$\rightarrow$S$\rightarrow$J transitions.
This is because  an F$\rightarrow$S transition in arrival traffic at the light signal leads to  MSP emergence
that
does not  necessarily cause the breakdown at the light signal. When MSP emergence
leads to the breakdown, there can be a long time-sequence of many
      F$\rightarrow$S transitions with MSP formation in each of the subsequent light signal cycles 
      before the breakdown (transition from
under- to over-saturated traffic) occurs.
These traffic phenomena that are characteristic ones for 
 the light signal do not occur at the on-ramp bottleneck.

 \subsection{Comparison of   traffic breakdown at light signal   within the frameworks of three-phase and
    two-phase traffic flow theories \label{2Phase_S}} 
    
    A two-phase  model  follows from the  three-phase model (Sec.~\ref{Model_S}) after 
removing   the description of driver behaviors  associated with three-phase theory~\cite{KernerBook,KernerBook2} --
  2D-region of synchronized flow  states (dashed region $S$ in Fig.~\ref{GW_model} (b))
  as well as a competition between the  speed adaptation 
  and over-acceleration effects  
  have been removed; this 
  is done through the use of   $G_{n}= 0$ and   $p_{\rm a}=p_{1}=p_{2}=0$  
  in the three-phase model (Appendix~\ref{App2}). As a result, 
  steady states of the two-phase model are related to a fundamental diagram (Fig.~\ref{GW_model} (c)).
 In the two-phase model traffic breakdown is governed 
 by a phase transition from free flow to the jam  (F$\rightarrow$J transition)~\cite{Reviews}.
 All   characteristics of  a wide moving jam
  in the three-phase and 
  two-phase models are {\it identical}, in particular, the flow rate in free flow   in the jam outflow  
  is equal to $q_{\rm out}=q_{\rm sat}=$ 1808 vehicles/h.
 Both   models exhibit the same and well-known
traffic behavior at light signal~\cite{Gartner}: (i) at small flow rates $\bar q_{\rm in}$ 
a vehicle queue   dissolves  fully during the green phase (under-saturated traffic), and (ii) at great enough $\bar q_{\rm in}$
the queue grows non-reversibly (over-saturated traffic) leading to
 traffic gridlock~\cite{Kerner2011_G}.
 
\begin{figure}
\begin{center}
\includegraphics*[width=8 cm]{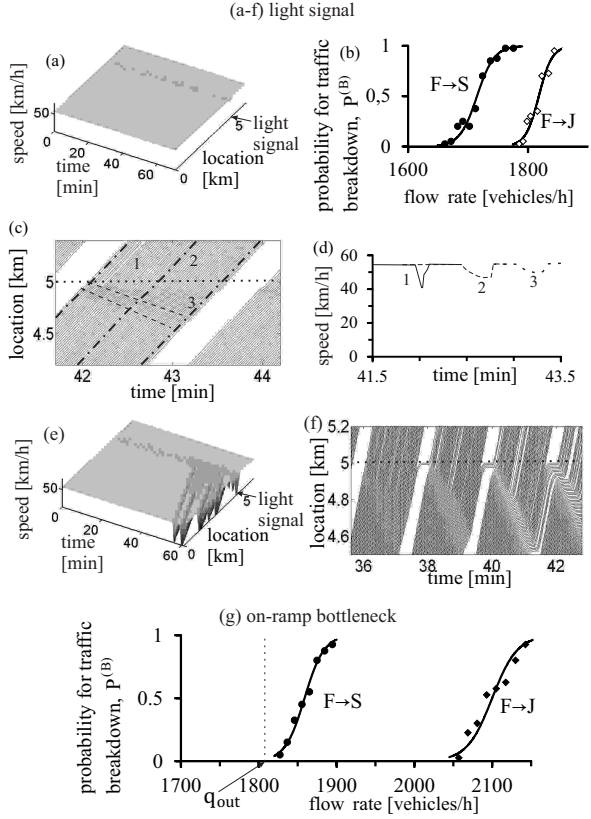}
\caption{Comparison of three-phase and two-phase models for simulations of traffic at
light signal (a--f) and at on-ramp bottleneck (g): (a) Speed in space and time 
simulated with two-phase model at the same model parameters as those in  Fig.~\ref{GW_3_2316}.
(b) Probability of breakdowns of green wave   as functions of   $\bar q_{\rm in}$ with three-phase model
(curve labeled by F$\rightarrow$S) and with two-phase model (curve labeled by F$\rightarrow$J).
(c, d) Vehicle trajectories (c) and associated  microscopic speeds  along vehicle
trajectories  1--3 (d) 
related to (c).
(e, f) Speed in space and time and trajectories for
green wave breakdown simulated with two-phase model at greater $\bar q_{\rm in}$ than that in (a, c, d).
(g) Probability for traffic breakdown at on-ramp bottleneck
in three-phase model (curve F$\rightarrow$S) and in two-phase model (curve F$\rightarrow$J).
 In (a, c--f), $q_{\rm GW}=$ 2316 (a, c, d) and 2446 vehicles/h (e, f).
  Other parameters in (a--f) are the same as those in Fig.~\ref{GW_3_2316_MSP}.
  In (g), $q_{\rm on}=$   300  vehicles/h.
  \label{FJ_no} } 
\end{center}
\end{figure}

Nevertheless, we have found that at the same flow rate $\bar q_{\rm in}=q_{\rm GW}(T_{\rm GW}/\vartheta)$
 and other model
 parameters as those used in  simulations   with   three-phase model shown in Fig.~\ref{GW_3_2316}
 in   {\it none} of simulation realizations made with two-phase model the
green wave  breakdown   can occur (Fig.~\ref{FJ_no} (a)).
 To understand this, note that at any given flow rate $\bar q_{\rm in}$ probability 
 of a sequence of F$\rightarrow$S$\rightarrow$J transitions occurring in
  three-phase model (curve labeled by F$\rightarrow$S in Fig.~\ref{FJ_no} (b)) 
 is considerably larger than probability of an F$\rightarrow$ J transition  
 that occurs in two-phase model (curve labeled by F$\rightarrow$J in Fig.~\ref{FJ_no} (b)).
 For this reason, although 
in two-phase model simulations there are also initial speed disturbances at the beginning
 of the green waves (Fig.~\ref{FJ_no} (c)),
 however, {\it no}  MSPs have emerged through these disturbances. This is because in contrast with
 in three-phase model, there is no synchronized flow in   two-phase model. As a result, in two-phase model
 the amplitude of the initial disturbances decreases during their propagation through the green waves (Fig.~\ref{FJ_no} (d)).

 The   result 
 that at any given flow rate probability of an F$\rightarrow$S  transition   is considerably greater
 than that of an F$\rightarrow$J  transition (Fig.~\ref{FJ_no} (b)) is a general one: This
 is also valid for a highway bottleneck as shown in Fig.~\ref{FJ_no} (g).

 Thus the green wave breakdown in two-phase model occurs  at
 considerably larger flow rates $\bar q_{\rm in}$ than those in three-phase model.
 At these large flow rates,  the initial disturbance with a considerably lower
 speed occurs causing long vehicle delays.  The subsequent  breakdown development is qualitatively similar to that
 found with three-phase models explained above (Figs.~\ref{FJ_no} (e, f)). 
 
 \begin{figure}
\begin{center}
\includegraphics*[width=9 cm]{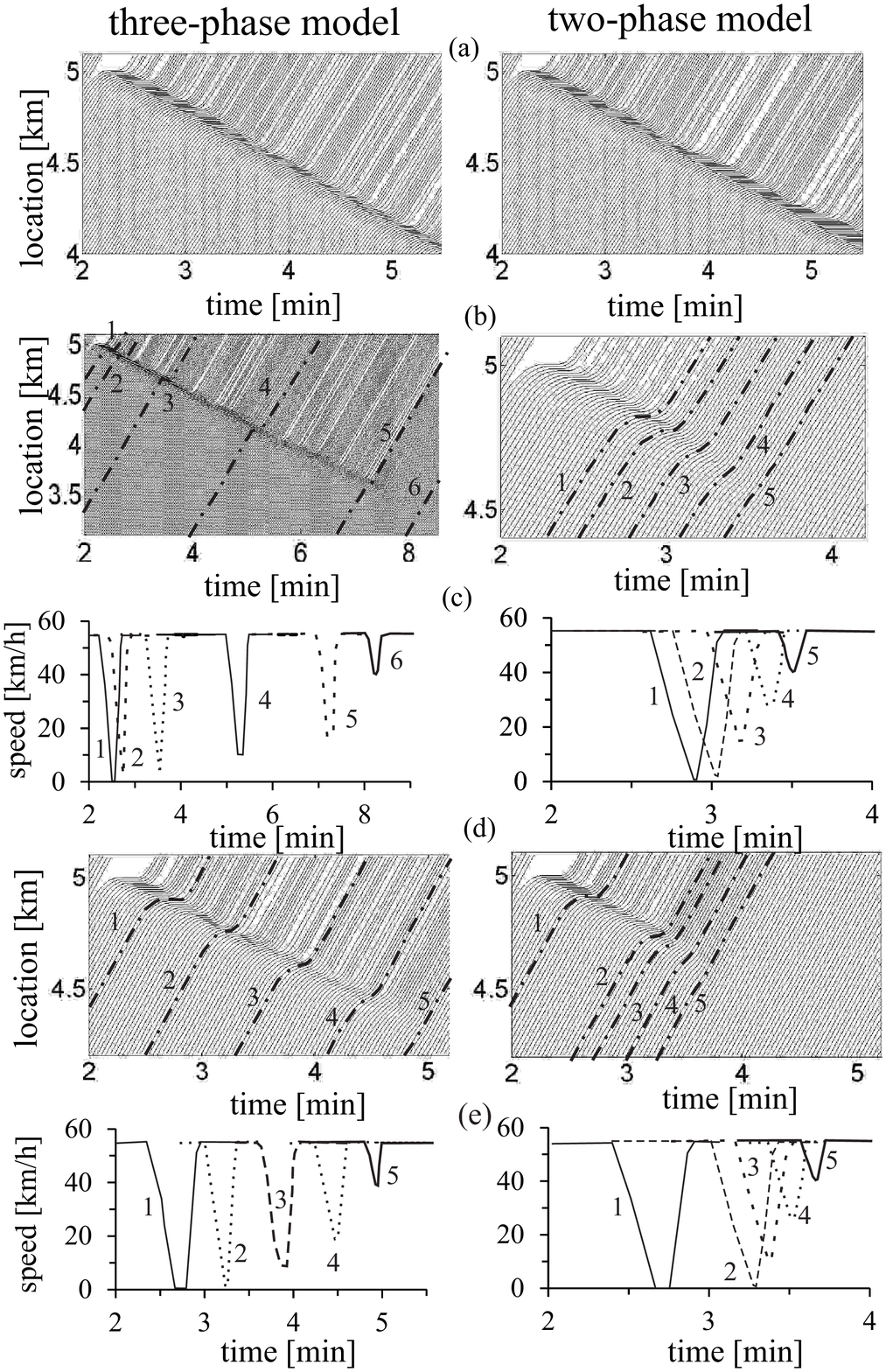}
\caption{Features of dissolution of wide moving jam (queue dissolution)
in three-phase model (left panel) and two-phase model (right panel):
(a, b, d) Vehicle trajectories for wide moving jam propagation without jam dissolution 
at $q_{\rm in}=q_{\rm out}$ (a) and under jam dissolution occurring at $q_{\rm in}<q_{\rm out}$ (b, d).
(c, e) Vehicle speed along trajectories related to (b, d), respectively.
Simulations on traffic flow on homogeneous road without light signal and other bottlenecks. To induced initial wide moving jam,
one of the vehicles   comes to a stop for 10 s; after this the vehicle accelerates in accordance with model rules
of vehicle motion. $q_{\rm in}=$ 1808 (a), 1800 (b, c) 1565 vehicles/h (d, e). 
  \label{MSP_dis} } 
\end{center}
\end{figure}

 The result that at a given flow rate $\bar q_{\rm in}$
 probability of a sequence of F$\rightarrow$S$\rightarrow$J transitions
 (three-phase model) is considerably larger than probability of an F$\rightarrow$J transition
 (two-phase model)
 remains also when $\bar q_{\rm in}$ is smaller than the threshold flow rate for 
 the MSP existence $q_{\rm th}$ in three-phase model (Sec.~\ref{WMJ_MSP_S}).
 This is   explained in Fig.~\ref{MSP_dis} through a consideration of
 the dissolution of a wide moving jam on a homogeneous road without light signals and
 other bottlenecks. In three-phase model, after the jam has dissolved a dissolving MSP occurs, which dissolves
 slowly (figures in left panel in Fig.~\ref{MSP_dis} (c, e)). This causes
 a much slower dissolution of a  local region of lower speed than this occurs in two-phase model
 (right panel in Fig.~\ref{MSP_dis} (c, e)) in which no synchronized flow
 can appear.

  \subsection{Conclusions \label{Conc_S}}

  1. There are very complex spatiotemporal self-organized traffic phenomena,
   which govern traffic behavior in city traffic, in particular, traffic capacity at the light signal.
  
  2. The delayed spontaneous  breakdown of a green wave
  is initiated by the emergence of an MSP within
  the green wave. The MSP causes   delays for vehicles that
   can randomly  lead to
   a  stop of one (or several) vehicle(s) moving at the end of the green wave. The discharge of a queue of the stopped vehicles causes
    an MSP with a lower synchronized flow speed, and so on.
     Long vehicle delays within an MSP result
    in a long queue build during the red phase. For one of the subsequent green waves, this queue
    cannot dissolve before   arrival of the following green wave. This causes traffic breakdown, i.e., the transition from under- to  
     over-saturated traffic at the light signal.

  3. There are the infinite number of capacities of traffic at the light signal, which are in a capacity range
  between a minimum capacity and maximum capacity. Each of the 
  capacities within the capacity range  gives the flow rate 
  at which  the breakdown can occur.

  4. The minimum capacity  is equal to the capacity of
    the classical  theory (Sec.~\ref{Int}).
    The maximum capacity determines the flow rate at which
    the random time-delayed breakdown occurs  spontaneously  during
  a given time interval with probability that is equal to 1.
  
  5. Within the capacity range, two capacity regions  separated by
  a threshold flow rate can be distinguished.
  In the first capacity region (between minimum capacity and threshold flow rate),
  an induced sequence of F$\rightarrow$S$\rightarrow$J transitions, i.e., the
  induced breakdown can occur only.
In the second capacity region (between   threshold flow rate and maximum capacity),
  a time-delayed spontaneous breakdown occurs  during
  a given time interval.

  6. At a given average arrival flow rate,
  both the maximum capacity and threshold flow rate   depend crucially on
  a time-dependence of this flow rate: The larger the number of vehicles  that arrive
  the light signal during the green phase of the light signal, the larger the maximum capacity and the larger the threshold flow rate.

 7. For time-functions of the flow rate studied, the largest maximum capacity and threshold flow rate are possible for a 
 hypothetical  green wave   in which
  all vehicles arrive  the light signal during the green phase only, i.e., when there is no an initial vehicle queue
  at the light signal.

     8.   The F$\rightarrow$S$\rightarrow$J transitions and   infinite number of capacities of traffic at the light signal
  can be well presented by a double Z-speed--flow-rate characteristic.
  
  9.   Probability of green wave breakdown as a function of light signal parameters can have a minimum in some flow rate range. 
  This can be used for green wave optimization with the BM principle.
  
   10. Green wave breakdown at a sequence of
   the light signals exhibits
   a complex spatiotemporal
dynamics of the breakdown  process associated with   MSPs
occurring upstream of different light signals. 

For a test of these and other conclusions, an empirical study 
of speed disturbances and  MSP emergence within a green wave should be made. To solve this problem,
measurements of  {\it single vehicle speed along the whole  green wave} are required that (for the author knowledge) are   not currently available.
Such measurements and their analysis will be an interesting task for a future study of the physics of traffic in a city.
 For engineering applications, an additional theoretical analysis of speed disturbances and MSP emergence
 within the green wave caused by left or
right turns and the width of the intersection can be  important.

{\bf Acknowledgments:}

  I thank German research and development  project $\lq\lq$UR:BAN" for support.
  I thank Sergey Klenov and Viktor Friesen for discussions  and Sergey Klenov for help in   simulations.

\appendix

\section{Definitions and symbols  \label{App_D}}

In   under-saturated  traffic at the light signal,
 all vehicles, which are waiting within a  queue  during the red phase, 
 can pass the signal during the green phase. 
 An opposite case occurs
 in   over-saturated  traffic and, therefore, the queue   grows.

Traffic   breakdown  at the light signal is the transition from under- to over-saturated traffic.
   Spontaneous  breakdown is   a random time-delayed breakdown.

Traffic   capacity  of the light signal  $C$ is   the average flow rate downstream of the light signal  
  $\bar q_{\rm LS}$
at which traffic   breakdown can occur at the light signal.

  Turning-in  traffic  refers
to traffic from the cross street that enters the lane on which the
green wave travels.

F -- free flow, S -- synchronized flow, J -- a moving queue at the light signal. An  
F$\rightarrow$S  transition is a local phase transition from free flow to synchronized flow occurring in arrival flow at the light signal.
The F$\rightarrow$S  transition leads  
   to the emergence of a moving synchronized flow pattern (MSP).
A sequence of F$\rightarrow$S$\rightarrow$J transitions  means the MSP emergence (F$\rightarrow$S  transition) with the subsequent
emergence of a moving queue (S$\rightarrow$J  transition) resulting in the breakdown  
 at the light signal.

$q_{\rm sat}$ is  the saturation flow rate,
i.e., the mean flow rate   from  a queue
 at the  light signal during the green phase when   vehicles discharge  
  to   the maximum free speed $v_{\rm free}$ ($q_{\rm sat}=$ 1808 vehicles/h under chosen model parameters).

$\vartheta=T_{\rm G}+T_{\rm Y}+T_{\rm R}$ is  the  cycle time of  the light signal.
 $T_{\rm G}$,
 $T_{\rm Y}$, and $T_{\rm R}$ are   durations of the
   green,   yellow,    and  red phases of the light signal, respectively.
  
 $x_{\rm b}$  and $x_{\rm LS}$ are coordinates of the 
road beginning  and isolated light signal, respectively.
In a light signal sequence, $\Delta x_{\rm LS}$ is a distance between the light signals that are
  at locations $x^{\rm (p)}_{\rm LS}, \ p=1, 2,..., P$, where $P$ is the number of the light signals.

 $T_{\rm GW}$  and $q_{\rm GW}$ are respectively the green wave   duration   
  and flow rate within green wave   given at location $x=x_{\rm b}$.

$\Delta T_{\rm b}$ is  a random time gap
 between the end of the red phase  
  and   beginning of the  green wave.
  $\Delta T_{\rm e}$ is a random time gap
between the end of    green wave and beginning of the red phase.
$\Delta T^{\rm (ideal)}_{\rm b}$ 
   and  $\Delta T^{\rm (ideal)}_{\rm e}$ are respectively 
   values   of   $\Delta T_{\rm b}$ 
   and $\Delta T_{\rm e}$  under hypothetical
 vehicle motion
   at  the speed $v_{\rm free}$.

   $T^{\rm (B)}$ is a random time delay of
   traffic breakdown  at the light signal.
   
  $T_{\rm ob}$ is a    time interval for   observing traffic (in all simulations $T_{\rm ob}=$ 60 min).
  
  $P^{\rm (B)}$ is breakdown probability   during the time interval $T_{\rm ob}$.
  
  $q_{\rm in}(t)$
 is the    rate of arrival traffic at the light signal
 that average value is $\bar q_{\rm in}=\vartheta^{-1}\int^{\rm \vartheta}_{0}{q_{\rm in}(t)}dt$. 
 
 $q_{\rm turn}$ is the flow rate in turning-in traffic.

 $q_{\rm LS}$
 is the    rate of    flow downstream of  light signal (in the light signal outflow)
 that average value is $\bar q_{\rm LS}=\vartheta^{-1}\int^{\rm \vartheta}_{0}{q_{\rm LS}(t)}dt$.
 
 $C_{\rm min}$ and $C_{\rm max}$ are respectively the minimum and maximum  
  traffic capacities at the light signal.
  
 $\bar q_{\rm in}=q^{\rm (B)}_{\rm th}$ is a threshold flow rate for spontaneous breakdown.

Variables and values  of a stochastic microscopic traffic flow model  used in simulations are explained
in   Appendix~\ref{App1}.

\section{Kerner-Klenov  three-phase microscopic traffic flow model for signal-lane 
   road     \label{App1}}
   
   Rules of vehicle motion in three-phase model, model functions, and model parameters used for simulations of self-organized traffic  
  are presented in Tables~\ref{table_FS_1}, \ref{table_FS_2}, and~\ref{table_FS_3}, respectively. 
  In the  model, discretized and dimensionless length (space coordinate), speed, and acceleration are used, which are
measured respectively in  discretization values 
$\delta x=$ 0.01 m,  $\delta v= 0.01 \  {\rm ms^{-1}}$, and $\delta a= 0.01 \  {\rm ms^{-2}}$; the value $\tau$
 is assumed to be
 dimensionless value $\tau=1$.
  With the
   exception of the mechanism of a stronger acceleration discussed in Sec.~\ref{Model_S},
  the  physics of the model has been explained in the book~\cite{KernerBook}.
    
   \begin{table}
\caption{Discrete version of stochastic three-phase traffic flow  model for single-lane road}
\label{table_FS_1}
\begin{center}
\begin{tabular}{|l|}
\hline
\multicolumn{1}{|c|}{
$v_{ n+1}=\max(0, \min({v_{\rm free}, \tilde v_{ n+1}+\xi_{ n}, v_{ n}+a_{\rm max}
\tau, v_{{\rm s},n} }))$,  
}\\
\multicolumn{1}{|c|}{
$x_{n+1}= x_{n}+v_{n+1}\tau$,
}\\
\multicolumn{1}{|c|}{
$\tilde v_{n+1}=\max(0, \min(v_{\rm free},  v_{{\rm s},n}, v_{{\rm c},n}))$,
} \\
\multicolumn{1}{|c|}{
$v_{{\rm c},n}=\left\{\begin{array}{ll}
v^{(1)}_{{\rm c},n} &  \textrm{at $\Delta v_{ n} + a_{ \ell,n}\tau < \Delta v_{\rm a}$,} \\
v^{(2)}_{{\rm c},n} &  \textrm{at $\Delta v_{ n} + a_{ \ell,n}\tau \geq \Delta v_{\rm a}$}, \\
\end{array} \right.$
} \\
\multicolumn{1}{|c|}{
$v^{(1)}_{{\rm c},n}=\left\{\begin{array}{ll}
v_{ n}+\Delta^{(1)}_{ n} &  \textrm{at $g_{n} \leq G_{ n}$,} \\
v_{ n}+a_{ n}\tau &  \textrm{at $g_{n}> G_{ n}$}, \\
\end{array} \right.$ 
} \\
\multicolumn{1}{|c|}{
$\Delta^{(1)}_{ n}=\max(-b_{ n}\tau, \min(a_{ n}\tau, \ v_{ \ell,n}-v_{ n})),$
} \\
\multicolumn{1}{|c|}{
$v^{(2)}_{{\rm c},n}=v_{ n}+\Delta^{(2)}_{ n},$ 
} \\
\multicolumn{1}{|c|}{
$\Delta^{(2)}_{ n}=k_{\rm a}a_{ n}\tau \max(0, \min(1, \ \gamma (g_{n}-v_{ n}\tau))),$
} \\
\multicolumn{1}{|c|}{
$g_{n}=x_{\ell, n}-x_{n}-d$, 
} \\
\multicolumn{1}{|c|}{
$\Delta v_{ n}=v_{ \ell,n}-v_{ n}$,
\  $a_{ \ell,n}=(v_{\ell, n}-v_{\ell, n-1})/\tau$,
} \\
\multicolumn{1}{|c|}{
$a_{\rm max}=\left\{\begin{array}{ll}
a &  \textrm{at $\Delta v_{ n} + a_{ \ell,n}\tau < \Delta v_{\rm a}$,} \\
k_{\rm a}a &  \textrm{at $\Delta v_{ n} + a_{ \ell,n}\tau \geq \Delta v_{\rm a}$}, \\
\end{array} \right.$
} \\
$v_{\rm free}$, $d$, $a$, $\Delta v_{\rm a}$, $k_{\rm a}$,  and $\gamma$ are constants;
 $\tau=1$; \\ $\ell$  marks the preceding vehicle. \\
\hline
\end{tabular}
\end{center}
\end{table}
\vspace{1cm}

 \begin{table}
\caption{Functions used in three phase model}
\label{table_FS_2}
\begin{center}
\begin{tabular}{|l|}
\hline
\multicolumn{1}{|c|}{Stochastic time delay of acceleration and
deceleration:} \\
\multicolumn{1}{|c|}{$a_{n}=a  \Theta (P_{\rm 0}-r_{\rm 1})$, \
$b_{n}=a  \Theta (P_{\rm 1}-r_{\rm 1})$,} \\
\multicolumn{1}{|c|}{
$P_{\rm 0}=\left\{
\begin{array}{ll}
p_{\rm 0} & \textrm{if $S_{ n} \neq 1$} \\
1 &  \textrm{if $S_{ n}= 1$},
\end{array} \right.
\quad
P_{\rm 1}=\left\{
\begin{array}{ll}
p_{\rm 1} & \textrm{if $S_{ n}\neq -1$} \\
p_{\rm 2} &  \textrm{if $S_{ n}= -1$},
\end{array} \right.$
}\\
\multicolumn{1}{|c|}{
$S_{ n+1}=\left\{
\begin{array}{ll}
-1 &  \textrm{if $\tilde v_{ n+1}< v_{ n}$} \\
1 &  \textrm{if $\tilde v_{ n+1}> v_{ n}$} \\
0 &  \textrm{if $\tilde v_{ n+1}= v_{ n}$},
\end{array} \right.$
}\\
$r_{1}={\rm rand}(0,1)$, $\Theta (z) =0$ at $z<0$ and $\Theta (z) =1$ at $z\geq 0$, \\
$p_{\rm 0}=p_{\rm 0}(v_{n})$,   $p_{\rm 2}=p_{\rm 2}(v_{n})$,
 $p_{\rm 1}$ is constant. \\
\multicolumn{1}{|c|}{Model speed fluctuations:} \\
\multicolumn{1}{|c|}{
$\xi_{ n}=\left\{
\begin{array}{ll}
\xi_{\rm a} &  \textrm{if  $S_{ n+1}=1$} \\
- \xi_{\rm b} &  \textrm{if $S_{ n+1}=-1$} \\
\xi^{(0)} &  \textrm{if  $S_{ n+1}=0$},
\end{array} \right.$
}\\
\multicolumn{1}{|c|}{$\xi_{\rm a}=a^{(\rm a)} \tau \Theta (p_{\rm a}-r)$, \
$\xi_{\rm b}=a^{(\rm b)} \tau \Theta (p_{\rm b}-r)$,} \\
\multicolumn{1}{|c|}{
$\xi^{(0)}=a^{(0)}\tau \left\{
\begin{array}{ll}
-1 &  \textrm{if $r\leq p^{(0)}$} \\
1 &  \textrm{if $p^{(0)}< r \leq 2p^{(0)}$ and $v_{n}>0$} \\
0 &  \textrm{otherwise},
\end{array} \right.$
}\\
$r={\rm rand}(0,1)$;
 $a^{(\rm b)}=a^{(\rm b)} (v_{n})$; \\  $p_{\rm a}$, $p_{\rm b}$, $p^{(0)}$, 
$a^{(\rm a)}$,  $a^{(0)}$ 
are constants.\\
\multicolumn{1}{|c|}{Synchronization gap $G_{n}$ and safe speed $v_{{\rm s},n}$:} \\
\multicolumn{1}{|c|}{
$G_{n}=G(v_{n}, v_{\ell,n})$,  
} \\
\multicolumn{1}{|c|}{
  $G(u, w)=\max(0,  \lfloor k\tau u+  a^{-1}\phi_{0}u(u-w) \rfloor),$
} \\
\multicolumn{1}{|c|}{
$v_{{\rm s},n}=
\min{(v^{\rm (safe)}_{ n},  g_{ n}/ \tau  + v^{\rm (a)}_{ \ell})}$, \ $v^{\rm (safe)}_{ n}=\lfloor v^{\rm (safe)} (g_{n}, \ v_{ \ell,n}) \rfloor$, 
} \\
 \multicolumn{1}{|c|}{
$v^{\rm (safe)} \tau_{\rm safe} + X_{\rm d}(v^{\rm (safe)}) = g_{n}+X_{\rm d}(v_{\ell, n})$,
} \\
 \multicolumn{1}{|c|}{
$X_{\rm d} (u)=b \tau^{2} \bigg(\alpha \beta+\frac{\alpha(\alpha-1)}{2}\bigg)$, $\alpha=\lfloor u/b\tau \rfloor$, $\beta=u/b\tau-\alpha$,
} \\
\multicolumn{1}{|c|}{
$v^{\rm (a)}_{\ell}=
\max(0, \min(v^{\rm (safe)}_{ \ell, n}, v_{ \ell,n},  g_{ \ell, n}/\tau)  -a\tau),$
} \\
 $\tau_{\rm safe}$
 is a safe time gap; 
$b$, $k>1$, and $\phi_{0}$ are constants; \\ $\lfloor z \rfloor$ denotes the  integer part of a real number $z$. \\
\hline
\end{tabular}
\end{center}
\end{table}
\vspace{1cm}

\begin{table}
\caption{Model parameters for three-phase model used in simulations}
\label{table_FS_3}
\begin{center}
\begin{tabular}{|l|}
\hline
$\tau_{\rm safe}   = \tau=1$, $d = 7.5 \  \rm m$,  \\
$v_{\rm free} = 15.278 \ {\rm ms^{-1}}$ ($55 \ \rm km/h$), \\
 $b = 1 \ {\rm ms^{-2}}$,  $a= 0.5 {\rm ms^{-2}}$, \\
$k=$ 3,  $\phi_{0}=1$,   $\Delta v_{\rm a}=2 \ {\rm ms^{-1}}$,
$k_{\rm a}=4$,  $\gamma=1$, \\
 $p_{b}=   0.1$,  $p_{\rm a}=0.03$, $p_{1}= 0.35$, $p^{(0)}= 0.005$, \\
$p_{\rm 2}(v_{n})=0.48+ 0.32\Theta{( v_{n}-v_{21})}$, \\
$p_{\rm 0}(v_{n})=0.667+ 0.083\min{(1, v_{n}/v_{01})}$, \\
$v_{01} = 6 \ {\rm ms^{-1}}$, $v_{21} = 7 \ {\rm ms^{-1}}$. \\
$a^{(\rm a)}= a$,  $a^{(0)}= 0.2a$, $a^{(\rm b)}(v_{n})=0.2a+$
\\ $+0.8a\max(0, \min(1, (v_{22}-v_{n})/\Delta v_{22})$, \\
 $v_{22} = 7 \ {\rm ms^{-1}}$,  $\Delta v_{22} = 2 \ {\rm ms^{-1}}$. \\
\hline
\end{tabular}
\end{center}
\end{table}
\vspace{1cm}

 \section{Two-phase microscopic
stochastic  traffic flow model for signal-lane 
   road     \label{App2}}
   
    Rules of vehicle motion in two-phase model, model functions, and model parameters used for simulations of self-organized traffic 
  are presented in Tables~\ref{table_FJ_1}  and~\ref{table_FJ_2}, respectively. 

 \begin{table}
\caption{Discrete version of stochastic two-phase traffic flow  model for single-lane road}
\label{table_FJ_1}
\begin{center}
\begin{tabular}{|l|}
\hline
\multicolumn{1}{|c|}{
$v_{ n+1}=\max(0, \min({v_{\rm free}, \tilde v_{ n+1}+\xi_{ n}, v_{ n}+a_{\rm max}
\tau, v_{{\rm s},n} }))$ 
}\\
\multicolumn{1}{|c|}{
 $x_{n+1}= x_{n}+v_{n+1}\tau$,
}\\
\multicolumn{1}{|c|}{
$\tilde v_{n+1}=\max(0, \min(v_{\rm free},  v_{{\rm s},n}, v_{{\rm c},n}))$,
} \\
\multicolumn{1}{|c|}{
$v_{{\rm c},n}=\left\{\begin{array}{ll}
v_{ n}+a_{ n}\tau  &  \textrm{at $\Delta v_{ n} + a_{ \ell,n}\tau < \Delta v_{\rm a}$,} \\
v_{ n}+\Delta^{(2)}_{ n} &  \textrm{at $\Delta v_{ n} + a_{ \ell,n}\tau \geq \Delta v_{\rm a}$}, \\
\end{array} \right.$
} \\
\multicolumn{1}{|c|}{
$\Delta^{(2)}_{ n}=k_{\rm a}a_{ n}\tau \max(0, \min(1, \ \gamma (g_{n}-v_{ n}\tau))),$
} \\
\multicolumn{1}{|c|}{
$g_{n}=x_{\ell, n}-x_{n}-d$, \ $\Delta v_{ n}=v_{ \ell,n}-v_{ n}$,
} \\
\multicolumn{1}{|c|}{
 $a_{ \ell,n}=(v_{\ell, n}-v_{\ell, n-1})/\tau$,
} \\
\multicolumn{1}{|c|}{
$a_{\rm max}=\left\{\begin{array}{ll}
a &  \textrm{at $\Delta v_{ n} + a_{ \ell,n}\tau < \Delta v_{\rm a}$,} \\
k_{\rm a}a &  \textrm{at $\Delta v_{ n} + a_{ \ell,n}\tau \geq \Delta v_{\rm a}$}, \\
\end{array} \right.$
} \\
$v_{\rm free}$, $d$, $a$, $\Delta v_{\rm a}$, $k_{\rm a}$, and $\gamma $ are constants;
 $\tau=1$; \\ $\ell$   marks the preceding vehicle. \\
\hline
\end{tabular}
\end{center}
\end{table}
\vspace{1cm}

  \begin{table}
\caption{Functions and parameters for two-phase model used in simulations}
\label{table_FJ_2}
\begin{center}
\begin{tabular}{|l|}
\hline
\multicolumn{1}{|c|}{Stochastic time delay of acceleration and
deceleration:} \\
\multicolumn{1}{|c|}{$a_{n}=a  \Theta (P_{\rm 0}-r_{\rm 1})$, \
$P_{\rm 0}=\left\{
\begin{array}{ll}
p_{\rm 0} & \textrm{if $S_{ n} \neq 1$} \\
1 &  \textrm{if $S_{ n}= 1$},
\end{array} \right.$
}\\
\multicolumn{1}{|c|}{
$S_{ n+1}=\left\{
\begin{array}{ll}
-1 &  \textrm{if $\tilde v_{ n+1}< v_{ n}$} \\
1 &  \textrm{if $\tilde v_{ n+1}> v_{ n}$} \\
0 &  \textrm{if $\tilde v_{ n+1}= v_{ n}$},
\end{array} \right.$
}\\
$r_{1}={\rm rand}(0,1)$, $\Theta (z) =0$ \\ at $z<0$ and $\Theta (z) =1$ at $z\geq 0$,
$p_{\rm 0}=p_{\rm 0}(v_{n})$. \\
\multicolumn{1}{|c|}{Model speed fluctuations:} \\
\multicolumn{1}{|c|}{
$\xi_{ n}=\left\{
\begin{array}{ll}
- \xi_{\rm b} &  \textrm{if $S_{ n+1}=-1$} \\
\xi^{(0)} &  \textrm{if  $S_{ n+1}=0$},
\end{array} \right.$
}\\
\multicolumn{1}{|c|}{
$\xi_{\rm b}=a^{(\rm b)} \tau \Theta (p_{\rm b}-r)$,} \\
\multicolumn{1}{|c|}{
$\xi^{(0)}=a^{(0)}\tau \left\{
\begin{array}{ll}
-1 &  \textrm{if $r\leq p^{(0)}$} \\
0 &  \textrm{otherwise},
\end{array} \right.$
}\\
$r={\rm rand}(0,1)$;
 $a^{(\rm b)}=a^{(\rm b)} (v_{n})$;   $p_{\rm b}$, $p^{(0)}$, 
 $a^{(0)}$ 
are constants.\\
Safe speed $v_{{\rm s},n}$ as well as the model parameters \\ and functions
$v_{\rm free}$, $d$, $a$,  $k_{\rm a}$, $\gamma$,  $p_{\rm 0}=p_{\rm 0}(v_{n})$, \\
$a^{(\rm b)}=a^{(\rm b)} (v_{n})$,   $p_{\rm b}$, $p^{(0)}$,  $a^{(0)}$ 
 are the same \\ as those in three phase model (Tables~\ref{table_FS_2}
 and~\ref{table_FS_3}). \\
\hline
\end{tabular}
\end{center}
\end{table}


\begin{thebibliography}{8.}
\bibitem{Ac_Fol}
  The increase in maximum acceleration 
    from standstill  does not influence on safety motion of following vehicles because
     vehicles accelerate after a time delay in acceleration whose mean
    value (about 1.74 s) remains the same as that in the former model~\cite{KernerBook}.
\bibitem{Long}
 
 In over-saturated traffic,
 moving queues can propagate many kilometers upstream during the time interval $T_{\rm ob}=$ 60 min.
 To avoid a   dependence of simulation results on upstream boundary conditions, 
 a long road section  $x_{\rm LS}-x_{\rm b}=$ 11 km is chosen for those model parameters at which speed disturbances that initiate
 the breakdown   occur  
  in a small neighborhood of the light signal location, i.e., disturbances occurring away of the light signal location(s)
  are negligibal for the breakdown. To make associated figures clear, 
 simulated road section $x_{\rm b} \leq x \leq  6$ km is not shown.
 This explains the value $x=x_{\rm LS}=$ 5 km shown in related figures. 
The exception is the case of green waves with $\Delta T^{\rm (ideal)}_{\rm b}=$ 8 s
   for which in figures  road co-ordinates  are the same as those in simulations.
\bibitem{qGW_cr} 
    Simulations show that if  the flow rate $q_{\rm GW}$ is slightly increased in comparison with used
in Fig.~\ref{GW_3_2316_MSP}, then MSP occurs
 in each of the cycles. This is because the initial disturbance  becomes  larger than
  the critical one
 in each of the cycles. Otherwise, when $q_{\rm GW}$ is slightly decreased, then
 probability of MSP occurrence decreases sharply.
\bibitem{Probability} 
 
   Probability $P^{\rm (B)}$ that traffic breakdown  at the light signal occurs spontaneously during
   the time interval $T_{\rm ob}$  
   is found through formula
$P^{\rm (B)}=n_{\rm b}/N_{\rm r}$,
where $N_{\rm r}$ is the number of different realizations (runs)~\cite{Realization},
 $n_{\rm b}$ is the number of runs in which traffic breakdown has occurred, therefore, $N_{\rm r} - n_{\rm b}$ is the number of runs in which
no breakdown has occurred.
 In all simulations, we have used $N_{\rm r}=40$.
\bibitem{Prob_ind}  
 We have also found that sometimes at a given value $M_{\rm turn}$   in some realizations (runs)~\cite{Realization}
  the breakdown does occur; however, in other realizations
   during a few following light signal cycles the initial disturbance decays fully, i.e., no breakdown occurs.
\bibitem{Fluc_Cap}
 
 A smoothing character of function
  $P^{\rm (B)}(\bar q_{\rm in})$ around $\bar q_{\rm in}= C_{\rm min}$ for the red wave (Fig.~\ref{Probability_FS} (g))
  is caused by the effect of model fluctuations
  on  queue discharge outflow.
\bibitem{zeta} 

For example, if   link inflow with the rate $q_{m}$
consists of $\lq\lq$fast", $\lq\lq$slow", and $\lq\lq$long"
vehicles with percentages $f_{\rm m}$, $s_{\rm m}$, and $l_{\rm m}$ (where $f_{\rm m}+s_{\rm m}+l_{\rm m}=100\%$),
then $\zeta_{m}=(f_{\rm m}, \ s_{\rm m}, \ l_{\rm m})$.
\bibitem{LS_cite}

When   the bottleneck $w$  is caused by the  light signal, then   $\alpha_{\rm w}=\alpha_{\rm w}(\vartheta_{\rm w}, \nu_{\rm w},
\beta_{\rm w}, \varphi_{\rm w}, \mu_{\rm w})$, where $\vartheta_{\rm w}$ is a cycle length of the light signal $w$,
 $\nu_{\rm w}$  is the offsets, i.e., the difference of the reference point of the respective local signal timing plan to the global (network-wide) 
 time scale reference point of the light signal phase begin, $\beta_{\rm w}$  defines the succession of the phases, 
 parameters $\varphi_{\rm w}$ and $\mu_{\rm w}$ define the start times and durations of the green and red phases for the light signal.
\end{thebibliography}
\end{document}